\documentclass[12pt, twoside]{article}
\usepackage{a4wide,amssymb,cite}
\usepackage{epsfig,axodraw}
\usepackage{color}

\newcommand{\be}{\begin{equation}}
\newcommand{\ee}{\end{equation}}
\newcommand{\bea}{\begin{eqnarray}}
\newcommand{\eea}{\end{eqnarray}}
\def\de{\partial}
\def\a{\alpha}
\def\b{\beta}
\def\g{\gamma}

\def\d{\delta}

\def\e{\eta}
\def\la{\lambda}

\def\k{\kappa}
\def\m{\mu}
\def\n{\nu}

\def\o{\omega}
\def\s{\sigma}

\def\f{\phi}

\def\th{\theta}
\def\Th{\Theta}
\def\de{\partial}
\def\z{\zeta}

\newcommand{\ie}{{\it i.e.,}\ }

\begin{document}

\begin{titlepage}

\rightline{July 2009}

\begin{centering}
\vspace{1cm}
{\Large  {\bf Codimension-2 brane inflation}}\\

\vspace{1.5cm}

{\bf Hyun Min Lee} $^{a,*}$ and {\bf Antonios Papazoglou} $^{b,**}$ \\
\vspace{.2in}

{\it $^a$Department of Physics \& Astronomy, McMaster University \\
Hamilton, Ontario L8S4M1, Canada. \\
\vspace{3mm}
$^{b}$Institute of Cosmology \& Gravitation, University of Portsmouth,\\
Portsmouth PO1 3FX, UK.} \\

\end{centering}

\vspace{1.5cm}

\begin{abstract}
\noindent
We consider a probe codimension-2 brane inflation scenario in a warped six-dimensional flux compactification. Our background model is Salam-Sezgin gauged supergravity with codimension-2 brane sources, which preserve ${\cal N}=1$ supersymmetry. The model has a modulus, which is stabilised by means of a cap regularisation of the codimension-2 singularities, with appropriate dilaton potentials on the ring interface of the caps with the bulk. We discuss the cosmological evolution of the world-volume of a probe codimension-2 brane when it moves along the radial direction of the internal space. In order to have slow-roll inflation, one needs the warping of the internal space to be weak, in contrast to the recent string inflation constructions with strong warping. We discuss the parameter range that the inflation is in agreement with the observationally inferred parameters and which furthermore is consistent with the probe brane approximation. We provide arguments pointing that the probe brane approximation is a good assumption if the probe brane is not exactly conical and show with a multi-brane solution that the mild warping needed for a slow-roll inflation is not spoiled by the probe brane.

\end{abstract}

\vskip 1cm

\vspace{2cm}
\begin{flushleft}

$^*~$E-mail: hminlee@mcmaster.ca \\
$^{**}~$E-mail: antonios.papazoglou@port.ac.uk

\end{flushleft}
\end{titlepage}

\section{Introduction}

Cosmological inflation\cite{inflation} has been the the most successful paradigm to solve the problems of standard big bang cosmology, as the horizon, the flatness, etc. Furthermore, it provides the seeds of structure in the Universe relating them to quantum fluctuations of the inflaton field, and so offering an interesting link between quantum scale physics and the macrocosm.  The origin of the inflaton field and its potential is a theoretical challenge and has been a subject of intense research for the past decades. One hopes that ideally this field can arise within some fundamental theory and its potential will be fixed by the latter. For example, there had been a lot of work of how inflation can be realised in supergravity or in string theory (for several reviews see \cite{reviews}). The D-brane inflation in strongly warped compactifications \cite{KKLT}
has drawn much attention recently. In that case the inflaton is identified as the separation between a brane and an anti-brane, as first considered in the unwarped case by \cite{DT}. Thus, it is crucial to achieve sufficiently weak inter-brane forces for a slow roll to occur.
On the other hand, in Dirac-Born-Infeld(DBI) inflation \cite{dbi,angularmotion1,dbihigher}, the brane position of a relativistically moving D-brane is considered as the inflaton.  These probe brane scenarios are distinguished from the one of the mirage cosmology \cite{mirage}, where the brane position is not an effective four-dimensional field interacting with the scale factor, but instead is directly related to the latter (see \cite{miragecosmo} for relevant models).

In the present paper we will follow a similar path by considering a six-dimensional supergravity model, the Salam-Sezgin model \cite{ss}, with two compact dimensions. The inflaton is the field corresponding to the position of a probe 3-brane, which moves in the internal compact space (for a different scenario of codimension-2 brane inflation see \cite{cline}). The warping of the internal space provides a potential for the inflaton. If the warping is small, one can see that the inflaton slow-rolls, with not so unnatural choice of parameters. The special characteristic of this model is that slow-roll inflation happens because of a {\it small} warping, in contrast to the usual string constructions where a {\it strong} warping is used for a red-shift in the effective brane tension \cite{KKLT}.

In more details, we consider the six-dimensional gauged supergravity of \cite{ss}, which has vacua featuring the mechanism of spontaneous compactification when a gauge field flux is turned on in the internal two-dimensional space. This construction and similar ones have been a very active theoretical laboratory for studying issues of codimension-2 branes as selftuning, their cosmology, local solutions etc., see for instance \cite{6dself,6dcosmo,GBind}.  We have supplemented this theory with magnetic couplings to codimension-2 branes or brane-localized Fayet-Iliopoulos (FI) terms, so that the brane action preserves ${\cal N}=1$ supersymmetry (SUSY) according to \cite{leepapa,lee}. This monopole background  has four-dimensional flat solutions, which are  ${\cal N}=1$ supersymmetric when there is no warping. If one allows for warping, the remaining SUSY is broken. The background is supported by two codimension-2 branes situated in the axis of symmetry of the internal space. Whether SUSY is completely broken or not, the system has an unfixed modulus due to a scaling symmetry of the equations of motion. Most recently, modulus stabilization and SUSY breaking have been discussed in the context of four-dimensional effective supergravity derived from six-dimensional gauged supergravity with supersymmetric codimension-2 branes \cite{choilee}.

Before proceeding with the study of probe brane inflation, we discuss the issue of stabilisation of the unfixed modulus, since it can influence the inflationary dynamics. We follow the prescription of \cite{uvcap} and consider the stabilisation of the modulus due to the regularisation dynamics of the  background codimension-2 branes. In particular, a regularisation which replaces these codimension-2 branes with caps joint to the rest of the bulk with ring-like codimension-1 branes \cite{regular,4dgravity}, provides a potential for the modulus, if the dilaton couplings to the brane do not respect the previously mentioned scaling symmetry. By further analysing the suggestion of \cite{uvcap} we see that one can find parameter space of the dilaton couplings to the branes for which the zero ring brane radius limit is purely conical, and therefore the bulk background solution remains unaffected. With this stabilisation procedure, the dilaton is naturally stabilised with the modulus mass being close to the compactification scale. A further advantage of this regularisation procedure is that, at least for the Einstein-Maxwell system\footnote{A generalisation of \cite{4dgravity} to the present supersymmetric model for scale invariant ring branes will give four dimensional scalar-tensor theory at large distances, with the scalar being the unfixed modulus. However, making the ring branes scale non-invariant, stabilises the modulus and decouples it at low energies.}, four dimensional gravity is reproduced on the brane at large distances \cite{4dgravity}, therefore our four dimensional effective treatment is justified. This is to be contrasted to the usual problem of obtaining four dimensional gravity on infinitesimally thin codimension-2 branes \cite{clown}.

The next step is to add a probe brane in the above background.  There are two types of probe branes that one may add in the bulk, codimension-1 and codimension-2. Regarding the codimension-1 branes, to have a proper four-dimensional effective theory on the brane, the radius of the 4-brane should either be small, thus resembling a codimension-2 brane, or should be centered around the axis of symmetry, \ie around one of the background codimension-2 branes. The second case, however, does not give rise to slow-roll inflation. Therefore, for our purposes the study of codimension-2 probes is sufficient.
For the slow-roll inflation of the codimension-2 probe brane, we need a mild warping which in turn requires the two background brane tensions to be almost equal.
In this case, we can match the bound on the spectral index, having a sufficient number of efoldings.
We also show that the COBE normalization of the density perturbation constrains the compactification scale to be of order $10^{13}$ GeV, thus the six-dimensional fundamental scale should be of order $10^{15}$ GeV. The inflation ends with the collision of the probe brane to one of the background branes. We provide a toy model describing a way to have graceful exit for the inflationary period, based on hybrid inflation \cite{hybrid}.

For the probe brane approximation to be justified, one needs to show that the backreaction of the probe brane to the background solution is negligible. There are two sources of backreactions possible: one is the backreaction on the volume modulus dynamics and the other is the backreaction on the warp factor. The first backreaction can be made negligible by taking the Hubble scale during the inflation to be much smaller than the modulus mass or the compactification scale. In turn, the tension of the probe brane should be small compared to the six-dimensional fundamental scale. Then, taking into account the COBE normalization, the scale of inflation (linked to the probe brane tension) should be smaller than the compactification scale. This gives the strictest bound on the warping of the four-dimensional space. Furthermore,
 a natural way to make the probe brane tension much smaller than the six-dimensional fundamental scale, is to make it slightly non-conical. We relate de-Sitter  non-conical probe  brane solutions to flat conical brane solutions and by studying explicit multi-brane flat solutions, we argue that the second source of backreaction of the probe brane, \ie the one to the warp factor, is negligible.

It is worth nothing that taking the probe brane slightly non-conical also solves an apparent paradox. Conical branes have a well known property to "hide" vacuum energy in a local deficit angle without curving their world-volume \cite{CLP}. This could pose a problem in our scenario of probe brane inflation, since inflation is driven by the potential energy of the probe. If this potential energy is "hidden", it will not gravitate as expected from the DBI action and our analysis would be incorrect. By taking the probe brane to be slightly non-conical, that is, the ring 4-brane that regularizes a non-conical probe brane, the above is avoided. Indeed, we show that the effective action in the presence of such a non-conical probe brane behaves as the naive DBI action constructed with a conical flat probe, for small departures of conicality.

The paper is organized as follows.
We first give a brief review on the general warped solution in the six-dimensional gauged supergravity and then discuss about the modulus stabilization in the presence of the dilaton potentials localized on the regularized background branes.
Next we derive the DBI action for probe branes and focus on the codimension-2 probe brane for a slow-roll inflation with a mild warping. Finally, we point out the constraint coming from the backreaction of the probe brane on the volume modulus and on the warping
and the conclusion is drawn.
Four appendices are added for providing the general non-conical warped flat solutions, the effect of a nonzero FI term
localized on the probe brane, the effective action approach to the probe brane potential and the general conical multi-brane flat solutions.

\section{Model setup}

Let us first make a short revision of the background that we will use for probe brane inflation. The theory that we will consider is the six-dimensional chiral gauged supergravity, also known as Salam-Sezgin supergravity \cite{ss}.  If codimension-2 branes are present in the extra dimensions at $y=y_i$, the action may be supersymmetrised  along the lines of  \cite{leepapa,lee}. Ignoring the Kalb-Ramond field and the hyperscalars in the bulk\footnote{
As far as hyperscalars and KR field don't couple to the probe brane, or if they get masses on the background branes, those bulk fields do not influence the dynamics of the probe-brane position as will be discussed in the later sections.
}, the bosonic action of the system\cite{lee} is
\bea
S={M_*^4 \over 2}\int d^6x \sqrt{-g_6}\left[ R_6-\frac{1}{4}(\partial_M\phi)^2-\frac{1}{2M^4_*}e^{\frac{1}{2}\phi}{\hat F}_{MN}{\hat F}^{MN}
-4g^2M^4_* e^{-\frac{1}{2}\phi}\right]+S_{\rm brane}~, \label{fullaction}
\eea
with
\be
S_{\rm brane}=-\int_{y=y_i} d^4x \sqrt{-g_4}\Big[e^{\frac{1}{2}\phi}(D_\mu Q_i)^2+2r_ig^2M^4_*|Q_i|^2+T_i+V_i(Q_i,\phi)\Big]~,
\ee
where the extra components of the modified gauge field strength are
\be
{\hat F}_{mn}=F_{mn}-\bigg(r_ig|Q_i|^2+\Big(\frac{2}{M^4_*}\Big)^{\frac{1}{2}}\xi_i\bigg)\epsilon_{mn}\frac{\delta^2(y-y_i)}{e_2}~.
\label{gaugefieldst}
\ee
In the above, $\epsilon_{mn}$ is the two-dimensional volume form, $e_2$ is the determinant of zweibein for extra dimensions and $M_*$ is the six-dimensional  fundamental scale. Moreover,  $T_i$ are the brane tensions, and $Q_i$ are brane scalar fields with covariant derivatives $D_\mu Q_i=(\partial_\mu-ir_i g A_\mu)Q_i$
and  brane scalar potentials $V_i(Q_i,\phi)$ coming from the brane $F$- and $D$-term \cite{lee}.
For a BPS brane, preserving ${\cal N}=1$ SUSY, the localized Fayet-Iliopoulos (FI) term is related to the brane tension as $\xi_i=\eta_i\frac{T_i}{4g}\Big(\frac{2}{M^4_*}\Big)^{\frac{1}{2}}$
with $\eta_i=\pm 1$ depending on the four-dimensional chirality of the brane SUSY\footnote{For the brane SUSY that is the same as the one preserved by the bulk flux, $\eta_i=+1$.}.

Assuming axial symmetry in the internal space and zero F- and D-term potentials $V_i(Q_i,\phi)$ for the brane scalar,
it has been found that the general warped solution with four-dimensional  Minkowski space\cite{gibbons,burgess03,leepapa} and conical branes, has $Q_i=0$ and takes the following form\footnote{From the warped solution with a dilaton constant set to zero\cite{gibbons,leepapa}, we recovered the dilaton constant $\phi_0$ by using the bulk scaling invariance under $g_{MN}\rightarrow e^{\frac{1}{2}\phi_0}g_{MN}$ and $\phi\rightarrow \phi+\phi_0$. The most general warped solution with conical branes but without assuming axial symmetry was found in Ref.~\cite{multibranes}.},
\bea
ds^2&=& e^{\frac{1}{2}\phi_0}\Big[W^2(r)
\eta_{\mu\nu}dx^\mu dx^\nu+A^2(r)(dr^2+B^2(r)d\theta^2)\Big]~, \nonumber  \\
{\hat F}_{mn}&=&q e^{-\frac{1}{2}\phi} W^{-4} \epsilon_{mn}~,  \label{wmetric}\\
\phi&=&\phi_0+4\ln W ~, \nonumber
\eea
with
\be
A={W \over f_0}~,  \  \  \  B={\lambda r \over W^4}~,
\ee
\be
W^4=\frac{f_1}{f_0}~, \ \ f_0=1+\frac{r^2}{r^2_0}~, \ \ \
f_1=1+\frac{r^2}{r^2_1}~,
\ee
where $\lambda$, $\phi_0$ and $q$ are constant parameters, and the two radii $r_0$, $r_1$ are given by
\be
r^2_0=\frac{1}{g^2M^4_*}~, \quad r^2_1=\frac{4M^4_*}{q^2}~.\label{radii}
\ee
The imbalance between the two radii $r_0$ and $r_1$ expresses the degree of warping of the internal space. It is worth noting that with this ansatz and zero F- and D-term potentials, one cannot find warped solutions with $\langle Q_i\rangle \neq 0$, thus the FI term is determined by the value of the brane tension. However, there exist warped solutions\cite{leepapa1} with $\langle Q_i\rangle \neq 0$ for non-trivial potentials $V_i(Q_i,\phi)$. We will not mention them explicitly here, but they will be important later on, when discussing the backreaction of the probe brane.

In the solution we noted above, the metric has two conical singularities,
one at $r=0$ and the other at $r=\infty$, which is at finite
proper distance from the former one. The singular terms coming from the deficit angles
at these singularities need to be compensated by brane tensions
$T_i$ ($i=1,2$) with the following matching conditions,
\bea
T_1&=&2\pi M^4_*(1-\lambda)~, \label{tension1}\\
T_2&=&2\pi M^4_*\Big(1-\lambda\frac{r^2_1}{r^2_0}\Big)~, \label{tension2}
\eea
if the angular coordinate has periodicity $2\pi$. After imposing the matching conditions, there is one parameter $\phi_0$ undetermined.
However, it has been shown that in the presence of the modulus potentials on the regularized
branes, one can have the modulus $\phi_0$ fixed at the four-dimensional Minkowski vacuum,
provided that the positions of the regularized branes are fixed by the dynamics on the branes\cite{uvcap}. We will explore this possibility in the next section.

On the other hand, the gauge field strength has two singular FI terms proportional
to the brane tensions at the conical singularities, so that they modify the gauge potentials
at the branes. Therefore, the FI terms modify the quantization condition of the gauge flux as
\be
{2 \la g M^4_* \over q} = n-\Big({2 \over M_*^4}\Big)^{\frac{1}{2}} {g \over 2\pi} (\xi_1 + \xi_2)~. \label{quant}
\ee
Consequently, using the brane matching conditions, (\ref{tension1}) and (\ref{tension2}),
we obtain the relation between the brane tensions as
\be
\Big(1-\frac{T_1}{2\pi M^4_*}\Big)\Big(1-\frac{T_2}{2\pi M^4_*}\Big)
= \bigg[n-\Big({2 \over M_*^4}\Big)^{\frac{1}{2}} {g \over 2\pi} (\xi_1 + \xi_2)\bigg]^2~. \label{quant2}
\ee
For the unwarped football solution, it is $T_1=T_2=2\pi M^4_*(1-\lambda)$ and $q=2gM^4_*$. For the BPS conditions  $\xi_1=\xi_2=\frac{T_1}{4g}\Big(\frac{2}{M^4_*}\Big)^{\frac{1}{2}}$, we see that eq.~(\ref{quant2}) is satisfied for $n=1$, independently of the brane tension. In other words, $\lambda$ is not quantized and arbitrary. This is in contrast with the non-supersymmetric brane action background \cite{quevedo,gibbons}. Moreover, for this unwarped football solution, four-dimensional ${\cal N}=1$ SUSY is preserved \cite{leepapa}.

\section{The modulus potential}

As mentioned previously, it is well known that the general warped solution has a massless modulus in the spectrum, corresponding to the arbitrary value of $\f_0$, which needs to be stabilised. In order to give mass to this modulus, we will follow the suggestion of \cite{uvcap} and replace the conical 3-branes with capped 4-branes, with non-trivial couplings to the dilaton.  The action for the ring branes with dilaton coupling is given by
\be
S_i=-\int d^5x\sqrt{-\g_i}\Big[V_i(\phi)+\frac{1}{2}U_i(\phi)(D_{\hat\mu}\sigma_i D^{\hat\mu}\sigma_i)\Big]~,
\ee
where $\g_{{\hat\mu}{\hat\nu}}^i$ is the induced metric on the branes, $V_i,U_i$ are dilaton couplings to the branes
and $\sigma_i$ is a brane Goldstone-like scalar fields.

The effective potential for the modulus can be found according to \cite{uvcap} after substituting the background solution to the total action and integrating the extra dimensions. In the presence of the dilaton potentials on the ring branes with ultraviolet (UV) caps,
the effective potential has been obtained in \cite{uvcap} as follows,
\be
S_{\rm pot}=\pi \int d^4x \sum_{i=1,2}\sqrt{-\g_i}\bigg[\Big(\frac{V_i}{2}+2\frac{dV_i}{d\phi}\Big)-\frac{1}{2}
\Big(\frac{U_i}{2}-2\frac{dU_i}{d\phi}\Big)g^{\theta\theta}(k_i-eA_\theta)^2\bigg]~, \label{vacenergy}
\ee
where $k_i$ is given from the solution for a brane sigma scalar, $\sigma_i=k_i\theta$. In the above, we will consider the constant $\f_0$ to be replaced with an $x$-dependent dynamical field. This will be an accurate approximation of the modulus field and (\ref{vacenergy}) will be a good estimate of its potential for energies lower than the compactification scale. As can be seen from the metric (\ref{wmetric}), the field $\f_0$ will have mixing with the four-dimensional curvature. We can eliminate this mixing by means of a four-dimensional coordinate redefinition, thus  going to the Einstein frame. In order to stress that the modulus which we will discuss is in the Einstein frame, we will name it $\psi(x)$ and re-write our metric and dilaton ansatz as
\bea
ds^2&=&e^{-\psi(x)}W^2(r){\tilde g}_{\mu\nu}(x)dx^\mu dx^\nu + e^{\psi(x)}A^2(r)(dr^2+B^2(r)d\theta^2)~,\label{metrickk}\\
\phi&=&\phi_B + 2\psi(x)~,
\eea
with $\phi_B=4 \ln W$ and $F_{r\theta}$ being $\psi$-independent (the determinant $\psi$-dependence cancels the $\psi$-dependence through $\f$ in (\ref{wmetric})). Then, the effective action becomes
 \be
S_{\rm eff}=\int d^4x \sqrt{-{\tilde g}} \left[\frac{1}{2}M^2_P R_4({\tilde g})-M^2_P(\partial_\mu\psi)^2-V_{\rm eff}\right]~,\label{modulusaction}
\ee
where the four-dimensional Planck scale is given by $M^2_P=\lambda \pi r^2_0 M^4_*$ and the effective potential is
\be
V_{\rm eff}=\pi  \sum_{i=1,2} e^{-\frac{3}{2}\psi}W^4AB\bigg[\Big(\frac{V_i}{2}+2\frac{dV_i}{d\phi}\Big)
-\frac{e^{-\psi}}{2A^2B^2}\Big(\frac{U_i}{2}-2\frac{dU_i}{d\phi}\Big)(k_i-eA_\theta)^2\bigg]~.\label{effmodulus}
\ee
The minimisation of this effective potential determines the value of $\psi=\psi_0$, which corresponds to the dilaton constant in the original solution (\ref{wmetric}).  It depends on the form of the dilaton functions that appear in the ring brane dynamics. It is evident from (\ref{modulusaction}) that once the modulus acquires a mass, four dimensional gravity will be reproduced along the lines of \cite{4dgravity}.

Finally, let us also remark on the relation between the effective modulus potential for the general axisymmetric warped solution with four-dimensional Minkowski space\footnote{See Appendix A for the general warped solution with general non-conical flat branes. These would be the zero radius limits of the ring branes that we discuss in this section.} and the ring-brane junction conditions discussed in \cite{uvcap}. The deviation from the conical limit is encoded in the parameter $\lambda_3$ appearing in (\ref{genwarp}) and one of the ring-brane junction conditions for the background solution [eq.(3.27) of Ref.~\cite{uvcap}] reads
\bea
\lambda_3 &=& W^4AB\Big[\Big(\frac{V_1}{2}+2\frac{dV_1}{d\phi}\Big)-\frac{e^{-\psi_0}}{2A^2B^2}\Big(\frac{U_1}{2}-2\frac{dU_1}{d\phi}\Big)
(k_1-eA_\theta)^2\Big] \nonumber \\
&=&- W^4AB\Big[\Big(\frac{V_2}{2}+2\frac{dV_2}{d\phi}\Big)-\frac{e^{-\psi_0}}{2A^2B^2}\Big(\frac{U_2}{2}-2\frac{dU_2}{d\phi}\Big)
(k_2-eA_\theta)^2\Big]~. \label{junctions}
\eea
We note here that we have corrected the sign error in the junction condition of the other ring-brane given in \cite{uvcap}.
Thus, by inserting the above conditions for the static solution with $\psi=\psi_0$ into the effective modulus potential (\ref{effmodulus}), we find that the effective vacuum energy vanishes, which is consistent with the flatness of the background solution.

\subsection{The case with exponential dilaton couplings}

As an explicit demonstration of the method used, let us consider the proposal of  \cite{uvcap} and take the brane dilaton couplings to have an exponential form as
\be
V_i=v_i e^{-\frac{1}{4}s_i\phi}~, \quad U_i=u_i e^{-\frac{1}{4}t_i\phi}~.
\ee
Then, for $s_i=1$ and $t_i=-1$, the dilaton couplings to the ring branes are scale-invariant and from (\ref{junctions}) we see that $\lambda_3=0$. In order to stabilise the modulus, one should break this scale invariance. Here we will demonstrate that $\lambda_3=0$ (\ie purely conical singularities for zero radius limit of the background ring branes) does not necessarily mean that the ring-branes should be scale invariant, something that was not apparent
in \cite{uvcap}.  We will see that it is possible to obtain solutions which break scale invariance, but however, in the zero radius limit remain in the conical class\footnote{There exist of course scale breaking ring-brane couplings, for which $\lambda_3\neq 0$. In this case, we can see from eq.~(\ref{junctions}) that  there is no conical limit for a vanishing ring-brane radius. In section 5, we will put a bound on $|\lambda_3|$ from the slow-roll conditions for a moving probe brane.}.

From eq.~(\ref{effmodulus}),
we obtain the effective potential for the specific couplings that we have considered as
\be
V_{\rm eff}=\sum_{i=1,2}\Big[C_i e^{-\frac{1}{2}(s_i+3)\psi}+D_i e^{-\frac{1}{2}(t_i+5)\psi}\Big]~, \label{moduluspot}
\ee
where the coefficients are given in terms of the ring brane parameters as
\bea
C_i&=& -\frac{1}{2}\pi W^4 AB v_i (1-s_i)e^{-\frac{1}{4}s_i\phi_B}~, \label{cs}\\
D_i&=& \frac{1}{4}\pi \frac{W^4}{AB} u_i(1+t_i)(k_i-eA_\theta)^2 e^{-\frac{1}{4}t_i\phi_B}~. \label{ds}
\eea
From the above, we again verify that if one has scale-invariant ring branes, \ie $s_i=1$ and $t_i=-1$, then $C_i=D_i=0$ and so there is no modulus potential generated from the ring branes.

Let us also see the relation between the effective brane tensions and the coefficients of the exponentials appearing
in the effective potential. After the modulus is stabilized to $\psi=\psi_0$, integrating the brane action over the angular direction of the ring branes gives the effective brane tensions of the limiting codimension-2 branes
\be
T_i=2\pi W^4AB e^{-\frac{3}{2}\psi_0}\Big(V_i+\frac{e^{-\psi_0}}{2A^2B^2}U_i(k_i-eA_\theta)^2\Big)~.
\ee
Therefore, defining ${\hat C}_i \equiv e^{-\frac{1}{2}(s_i+3)\psi_0}C_i$ and ${\hat D}_i \equiv e^{-\frac{1}{2}(t_i+5)\psi_0}D_i$, the effective brane tensions are given by
\be
T_i=-\frac{4{\hat C}_i}{1-s_i}+\frac{4{\hat D}_i}{1+t_i}~, \quad i=1,2. \label{efftension}
\ee

\subsection{The minimization of the effective potential}

Let us see now explicitly how it is possible to obtain non-scale invariant ring branes which provide a stabilisation mechanism for the modulus and still remain in the conical class, \ie $\la_3=0$, in the thin ring brane limit. From eq.~(\ref{junctions}) for the static solution with $\psi=\psi_0$,
we obtain the relations between the coefficients in the effective potential (\ref{moduluspot})
as following
\be
{\hat C}_1+{\hat D}_1=-({\hat C}_2+{\hat D}_2)=0 ~.\label{juncsimple}
\ee
On the other hand, the extremization of the effective potential (\ref{moduluspot})
gives rise to another relation between the same coefficients
\be
\sum_{i=1,2} \Big({\hat C}_i(s_i+3)+{\hat D}_i(t_i+5)\Big)=0~.
\ee
Then, the above system of three equations with four unknowns can be solved in terms of $\hat{C}_1$. {\it E.g.,} for $\hat{C}_2$ one obtains
\be
{\hat C}_2=-{\hat C}_1\frac{s_1-t_1-2}{s_2-t_2-2}~. \label{c2}
\ee
The minimum of the modulus $\psi_0$ can then be found as following
\be
e^{-\frac{1}{2}(s_2-s_1)\psi_0}=-\frac{s_1-t_1-2}{s_2-t_2-2}\Big(\frac{C_1}{C_2}\Big)~.\label{minmodulus}
\ee
Substituting in (\ref{moduluspot}) the coefficients $\hat{C}_2$, $\hat{D}_1$, $\hat{D}_2$, we obtain the effective potential for the system as
\bea
V_{\rm eff}&=&{\hat C}_1\bigg[e^{-\frac{1}{2}(s_1+3)(\psi-\psi_0)}-e^{-\frac{1}{2}(t_1+5)(\psi-\psi_0)} \nonumber \\
&&\quad\quad-\frac{s_1-t_1-2}{s_2-t_2-2}\Big(e^{-\frac{1}{2}(s_2+3)(\psi-\psi_0)}-e^{-\frac{1}{2}(t_2+5)(\psi-\psi_0)}\Big)\bigg]~.
\eea
Taking into account the normalization of the modulus kinetic term in eq.~(\ref{modulusaction}), the quadratic in $(\psi-\psi_0)$ piece of the above potential provides us with the mass of the modulus at the minimum as
\be
m^2_{\psi}=\frac{1}{4M^2_P}\sum_{i=1,2}\Big({\hat C}_i(s_i+3)^2+{\hat D}_i(t_i+5)^2\Big)~.\label{modulusmass0}
\ee
Using eqs.~(\ref{juncsimple}) and (\ref{c2}), we can then express the modulus mass (\ref{modulusmass0}) in a simpler form
\be
m^2_{\psi}=\frac{1}{2M^2_P}{\hat C}_1(s_1-t_1-2)(s_1+t_1-s_2-t_2)~.\label{modulusmass}
\ee
Therefore, for $s_i,t_i$ of order 1, the modulus mass is of order ${\hat C}_1$.
 Assuming that both $u_i>0$ and $v_i>0$, one can see that there is no parameter space where the modulus is stabilised ($m_\psi^2 >0$). Keeping $u_i>0$ in order not to have ghost kinetic terms for $\s_i$, we can assume that only $v_1>0$, but $v_2<0$. From eq.~(\ref{cs}), the positive modulus squared mass requires
\be
(1-s_1)(s_1-t_1-2)(s_1+t_1-s_2-t_2)<0~.
\ee

We can distinguish four cases of possible parameter space where the modulus is stabilised
\bea
(I)&:&\,\, s_1<1, \,\, t_1>-1; \,\, s_2<1, \,\, t_2<-1; \,\, s_2-t_2-2<0; \\
(II)&:&\,\, s_1<1, \,\, t_1>-1; \,\, s_2>1, \,\, t_2>-1; \,\, s_2-t_2-2>0; \\
(III)&:&\,\, s_1>1, \,\, t_1<-1; \,\, s_2<1, \,\, t_2<-1; \,\, s_2-t_2-2<0; \\
(IV)&:&\,\, s_1>1, \,\, t_1<-1; \,\, s_2>1, \,\, t_2>-1; \,\, s_2-t_2-2>0,
\eea
where in all cases one should also satisfy $s_1+t_1 > s_2 + t_2$.

For $\lambda_3=0$, using eq.~(\ref{juncsimple}), the effective brane tensions (\ref{efftension}) become
\bea
T_1&=&-4{\hat C}_1\frac{(2-s_1+t_1)}{(1-s_1)(1+t_1)}~, \label{ringt1}\\
T_2&=&-4{\hat C}_2\frac{(2-s_2+t_2)}{(1-s_2)(1+t_2)}~. \label{ringt2}
\eea

For the above parameter space we see that for the cases II, III, IV, both branes have the same sign of tension $T_i>0~(i=1,2)$, but for the case I they have opposite sign, $T_1>0$ and $T_2<0$. Since we discuss solutions with small warping, we should have
both background 3-branes of the same tension sign. The only possibility where the case I can be allowed is if $T_1 \to 0^+$, $T_2 \to 0^-$, but this case is not interesting, since the modulus becomes very light, not allowing the inflation scenario that we'll discuss in the following sections.

For illustration, let us take the dilaton couplings as $(s_2,t_2)=(3,0)$ and $(s_1,t_1)=(\frac{1}{2}(1-\varepsilon),3)$. This example
falls into the case II noted above. In this case, from eqs.~(\ref{c2}), (\ref{ringt1}) and (\ref{ringt2}),
we obtain the ratio of brane tensions
as
\be
\frac{T_2}{T_1}=\frac{1}{2}(1-s_1)(1+t_1)=1+\varepsilon~.
\ee
Therefore, the brane tensions can be almost equal for $\varepsilon\ll 1$, which is the case that the probe brane inflation could be realized as we will see later. From eq.~(\ref{modulusmass}) with the minimization condition (\ref{minmodulus}), for $\varepsilon\ll 1$, we get the modulus minimum and the modulus mass, respectively, as
\bea
e^{-\frac{5}{4}\psi_0}&\simeq&\frac{9C_1}{2C_2}~, \label{egmodulusvev} \\
m^2_\psi&\simeq& \frac{9}{8M^2_P}\Big(\frac{9C_1}{2C_2}\Big)^{\frac{7}{5}}C_1~. \label{egmodulusmass}
\eea

\section{The probe brane action}

We will now introduce probe branes in the above background, whose motion will induce a cosmological evolution in their worldvolumes. The usual action for a probe brane is the DBI action supplemented by appropriate Chern-Simons (CS) terms
\be
S_{probe}=- T_p \int_{probe} f(\f) \sqrt{-\g_{ind}} + Q_p \int_{probe} C_{p+1}+\xi_p \int_{probe} F_{p+1}+\cdots~ \label{dbifull},
\ee
where $f(\f)$ is a suitable function of the dilaton, $\g_{ind}$ the induced metric on the probe brane, $C_{p+1}$ the pull back of a bulk form field and $F_{p+1}$ the pull back of a bulk field strength. The ellipsis denotes terms having to do with the coupling of brane gauge sectors to bulk fields. In the absence of non-zero background field strengths in the brane sectors, we will ignore the ellipsis. Let us now comment in more details for the various terms of (\ref{dbifull}) for the specific case of a codimension-2 probe brane. For such a brane, there is no $Q_p$-type coupling to a bulk form field, since that would have to be a 4-form bulk field or a dual bulk scalar field, both absent in the theory that we study. Compared to the D-brane action in string theory, this corresponds to the case that the Ramond-Ramond ($RR$) field is not present or the net $RR$ charge of the brane vanishes in six-dimensional effective supergravity. Instead, the field strength of the bulk $U(1)_R$ couples to the codimension-2 brane as a localized FI term ($\xi_p$-type coupling in (\ref{dbifull})), through the dual 4-form field strength $F_4=e^{\frac{1}{2}\phi}*F_2$. This localized FI term also carries the dilaton coupling as seen from the expansion of the modified gauge kinetic term in eq.~(\ref{fullaction}). For a SUSY codimension-2 brane with nonzero tension, on top of the linear FI term, there is a delta function squared term too \cite{leepapa,lee}. In the following, we will consider non-supersymmetric probe branes, in which case the localized FI term can be made negligible by assuming that the scale of $\xi_p$ is much smaller than the brane tension.

In the absence of the background branes, the bulk equations of motion are scale invariant. Under the scaling
\be
g_{MN} \to e^{\f_0/2} g_{MN} \, \, \, , \, \, \, \f \to \f + \f_0~,
\ee
the bulk action in (\ref{fullaction}) transforms as $S \to e^{\f_0} S$.  As explained in the previous section, a large breaking of this scale invariance on the capped background 4-branes is crucial for the stabilisation of the extra dimensional space. On the other hand, it is favourable that the probe branes that we are adding to the system approximately respect this symmetry. This will become clearer when we discuss the   validity of the DBI action and backreaction issues in Sec. 6 (and Appendix C in more details). There, we  will consider a small violation of the scaling symmetry for a probe ring brane with non-conical small radius limit and show that this naturally makes the effective probe brane tension much smaller than the overall scale of the modulus potential as required for the probe brane approximation. Since the violation of the scaling symmetry on the probe brane will be eventually taken to be small, we will neglect it in this and the next sections where we discuss the probe brane dynamics.  In this case, we can determine the precise form of the dilaton coupling to them. So, parameterising the coupling of the  probe 3-branes and 4-branes, respectively, as an exponential like
\be
S_{\rm probe}=-T_3 \int d^4x ~ e^{\z_3 \f} \sqrt{-\g_{\m\n}} \, \, \, , \, \, \, S_{\rm probe}=-T_4 \int d^5x ~e^{\z_4 \f} \sqrt{-\g_{ab}}~.
\ee
we see that they scale as the bulk action when $\z_3=0$, $\z_4=-1/4$.

\subsection{Probe 3-brane}

Let us first consider a codimension-2 probe brane with tension $T_3$, moving in the previous general warped background. The embedding coordinates of the probe brane are given by $X^M$. Suppose that the probe brane coordinate is given by $X^{M}=[x^\m,y_p]$ with $y_p=[R(t),\Th(t)]$ where $x^0=t$. Then, the induced metric on the probe brane is given by
\be
\g_{\m\n}= g_{MN} {\de X^M \over \de x^\m}{\de X^N \over \de x^\n}~.
\ee
For the metric (\ref{metrickk}), taking into account the modulus dependence, the 4D effective action becomes
\be
S_{\rm eff}=\int d^4x \sqrt{-{\tilde g}} \left[\frac{1}{2}M^2_P R_4({\tilde g})-M^2_P(\partial_\mu\psi)^2-V_{\rm eff}\right]+S_{\rm probe}~,
\ee
where the worldvolume action for the probe brane\footnote{Here we assume that the background is homogeneous, so that the brane position depends on time only. However, when the background is non-homogeneous, there will be extra terms in the metric determinant in (\ref{dbi3brane}) quartic in derivatives of the brane position. Those terms can give an important contribution to isocurvature perturbations and non-Gaussianities \cite{dbihigher}. We thank S\'ebastien Renaux-Petel for pointing this out.} is given by
\bea
S_{\rm probe}&=&-T_3 \int d^4x \sqrt{-\det \g_{\m\n}}\\&=&
-T_3 \int d^4x \sqrt{-{\tilde g}}e^{-2\psi}W^4(R(t))\sqrt{1+e^{2\psi}W^{-2}{\tilde g}_{mn}{\tilde g}^{\mu\nu}\partial_\mu X^m \partial_\nu X^n}~. \label{dbi3brane}
\eea
For small velocities, \ie $v^2\equiv W^2(r=R(t))|{\tilde g}_{mn}{\tilde g}^{00}{\dot X}^m{\dot X}^n|\ll 1$,
the worldvolume action gives
\be
S_{\rm probe}
\approx T_3
\int d^4x \sqrt{-{\tilde g}}\left[ \frac{1}{2}v^2 - W^4 e^{-2\psi} \right]~.
\ee
We, therefore, see that there is no kinetic mixing between the probe brane and the volume modulus,
while the probe brane potential depends on the volume modulus.
Here, the kinetic term for the radial motion can be written in a canonical form for
\be
d\chi = \sqrt{T_3}A(R) W(R)dR~, \label{canfield}
\ee
with $A=\frac{W(R)}{f_0(R)}$.
Thus, the approximate probe brane potential is given by
\be
V_{\rm probe}= T_3\,W^4(R)
e^{-2\psi}=T_3\,\left(\frac{1+\frac{R^2}{r^2_1}}{1+\frac{R^2}{r^2_0}}\right)e^{-2\psi}~.
\label{pot}
\ee
The total potential for the effective dynamics is
\be
V_{\rm tot}= V_{\rm eff}+V_{\rm probe}~. \label{totalpot}
\ee
There is a delicate question about whether the probe brane potential describes in a good way the effective four-dimensional dynamics of the system with moving probe codimension-2 brane. We will discuss this in Sec. 6.1 and argue that indeed this is true, for a slight deviation from a conical probe brane limit. We will, therefore, continue studying (\ref{totalpot}) for the effective dynamics.

If we assume that $T_3 \ll V_{\rm modulus}\sim M^4_*$, then the probe brane contribution to the total potential is negligible and
the minimum of $\psi$ is determined from the extremisation of $V_{\rm eff}$ alone. Then, once $\psi$ is fixed by
this minimisation,  the effective potential depends only on the radial direction.
The potential has a minimum at
$R=0$ for $r_0>r_1$ and $R=\infty$ for $r_0<r_1$.
In both cases, the probe brane tends to move to the background codimension-2 branes
located at the poles of the compactification.

Let us note that as mentioned in the beginning of the section, we are mainly interested in the non-supersymmetric probe branes. There is in principle a non-zero localised FI term  of the probe brane (contributing to the CS term in the brane effective action) whose effect can be under control as shown in Appendix B. For a supersymmetric brane, however, there is a relation between $\xi_3$ and $T_3$ [$\xi_3= \pm T_3/ (4g)(2/M^4_*)^{1/2}$] and therefore the analysis will break down. However, even then, one can use the vacuum expectation value (VEV) of a probe brane scalar $Q_3$ to set the effective $\xi_{3,\rm eff}\equiv \xi_3+r_3g|\langle Q_3\rangle|^2(M^4_*/2)^{1/2}$ to zero, while the effective brane tension $T_{\rm eff}=T_3+2r_3 g^2 M^4_* |\langle Q_3\rangle|^2+V_3(\langle Q_3\rangle)$ is nonzero. Then the subsequent analysis will hold if we substitute $T_3$ with $T_{3,\rm eff}$.

\subsection{Probe 4-brane}

The general way to embed a probe 4-brane in the internal space is $y_p=[R(t,\o),\Th(t,\o)]$, where $\o$ is the intrinsic angular coordinate of the 4-brane. There are two distinct cases that we are going to investigate. The first one, is a probe 4-brane that (symmetrically) encircles one of the background ring branes at the poles of the compactification. The second case, is a probe 4-brane that one obtains from the ring-regularisation of a probe 3-brane.

In the first case, because of the present symmetry we have $\o=\th$ and the 4-brane embedding is $y_p=[R(t),\th]$. The radius of the 4-brane is $R(t)$ and is changing because of the brane radial motion. The probe brane action is given by
\bea
S_{\rm probe}&=&-T_4 \int d^4x d\o~ e^{-\f/4}  \sqrt{-\det \g_{ab}}\\&=&
-2\pi T_4 \int d^4x \sqrt{-{\tilde g}}~e^{-2\psi} AB W^3 \sqrt{1+e^{2\psi}A^2 W^{-2} {\tilde g}^{\mu\nu}\partial_\mu R \partial_\nu R}~.
\eea

For low velocities, the probe potential is given by
\be
V_{\rm probe}= 2\pi T_4\,A(R)B(R)W^3(R) e^{-2\psi}~,
\ee
and the canonical field of the radial motion is
\be
d\chi= \sqrt{2 \pi T_4}\sqrt{A(R) B(R) W(R)}dR~.
\ee

In the second case, one has to do with a 4-brane whose average radius $R_0$ is very small compared with the average curvature radius of the internal space $r_0$, \ie $R_0 \ll r_0$. Then, the embedding resembles the one of the 3-brane $y_p \approx [R(t),\Th(t)]$. The limit, however, has to be taken carefully. The probe brane action will be given by
\bea
S_{\rm probe}&=&-T_4 \int d^4x d\o~ e^{-\f/4} \sqrt{-\det \g_{\m\n}}\\&\approx&
-2\pi R_0 T_4 \int d^4x \sqrt{-{\tilde g}}~e^{-2\psi} W^4 \sqrt{1+e^{2\psi}W^{-2}{\tilde g}_{mn}{\tilde g}^{\mu\nu}\partial_\mu X^m \partial_\nu X^n}+ {\cal O}(R_0^2)~,~~~~~~
\eea
where the approximation in the second line is based on the expectation that there should be a smooth limit between the dynamics of the probe 4-brane and the probe 3-brane to which the former corresponds in the limit of zero radius. We will give further  arguments for this approximation in the following.

Therefore, for low velocities, the probe potential is similar to the 3-brane potential
\be
V_{\rm probe} \approx 2\pi R_0 T_4\,W^4(R) e^{-2\psi} + {\cal O}(R_0^2)~.
\ee
In the strict 3-brane limit $T_4 \to \infty$, $R_0 \to 0$, with $T_3= 2\pi R_0 T_4 \to {\rm const.}$ Thus, this limit gives very similar results with the probe 3-brane modulo $R_0$-corrections. Therefore, we will not consider further this example and regarding the probe 4-brane, we will only check the first symmetric case.

Concerning the approximation, we will consider as a toy example a circular ring brane of radius $R_0$ moving in flat {\it unwarped} background with flat internal space, \ie the metric is simply
\be
ds^2=\e_{\m\n}dx^\m dx^\n + dr^2 + r^2 d\th^2~.
\ee
Then the description of the 4-brane embedding is tractable analytically and we can see how for small probe 4-brane radius we obtain the probe 3-brane result. Let the center of mass $x_0(t)$ of the ring brane move along the $x$-axis \ie with $\th=0$. Then if $\o$ is the intrinsic angular coordinate of the 3-brane, the embedding reads
\bea
R(t,\o)&=& \sqrt{R_0^2 + x^2_0(t) + 2 R_0 x_0(t) \cos \o }~,
 \\ \Th(t,\o)&=& \tan^{-1} \left({R_0 \sin \o \over x_0(t)+R_0 \cos \o} \right)~.
\eea
The action of this brane (with zero dilaton coupling for simplicity) is
\be
S_{\rm probe}=-T_4 \int d^4x d\o~ R \sqrt{(1-\dot{R}^2- R^2 \dot{\Th}^2)(\Th'^2 + R'^2/R^2)}~,
\ee
where $' \equiv d/d\o$. Substituting the embedding of the brane in the above and expanding in powers of $R_0$, we see that the leading term is
\be
S_{\rm probe} \approx -2\pi R_0 T_4 \int d^4x  \sqrt{1-\dot{x}_0^2} + {\cal O}(R_0^2)~,
\ee
which reproduces the action for a 3-brane moving as the center of mass of the ring brane with tension $T_3=2\pi R_0 T_4$. If the internal space is curved and the four dimensional space warped, the equivalence is expected to be harder to prove, but at the end it is expected to hold.

\section{Inflation in the background with a mild warping}

\begin{figure}[t]
\begin{center}
\begin{picture}(200,160)(0,0)

\Vertex(94,46){2.5}
\LongArrow(94,46)(120,55)
\Text(80,40)[c]{(A)}

\BCirc(94,96){5.5}
\LongArrow(94,96)(120,85)
\Text(80,109)[c]{(B)}

\Oval(170,71)(18,9)(0)
\LongArrow(170,71)(200,71)
\Text(145,71)[c]{(C)}

\epsfig{file=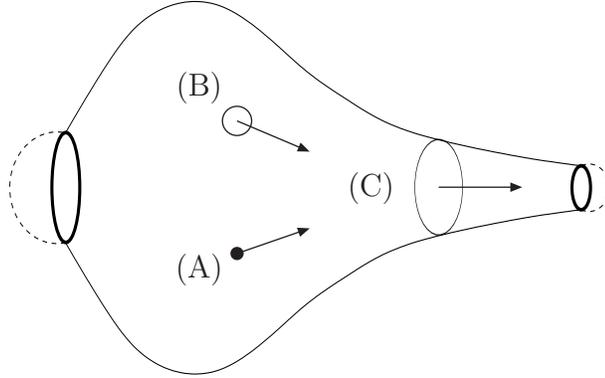,width=5cm,height=8cm,angle=90}

\end{picture}
\caption{The internal space with the capped ring background branes and the various probe branes: (A) a 3-brane, (B) a 4-brane of small radius resembling a 3-brane, (C) a 4-brane symmetrically encircling one background brane.  If the warping is very weak, the motion of the branes of type (A), (B), gives rise to slow-roll inflation, which ends when the probe hits one of the background branes.}
\label{internalfig}
\end{center}
\end{figure}

For the general warped background, it has been known that SUSY is broken completely.
However, for a constant warp factor with equal brane tensions, \ie $r_0=r_1$ or $T_1=T_2$,
four-dimensional ${\cal N}=1$ SUSY is known to be preserved
due to the localized FI terms \cite{leepapa,lee}.
When we regard the position of the probe brane as being an inflaton,
we will be led to assume a mild warping with $r_0 \sim r_1$ for which SUSY is slightly broken in the background.
The probe brane is supposed to start rolling from the background brane situated at one pole and end when the probe brane hits the background brane situated at the other pole. The above interesting result is for codimension-2 branes, or for ring codimension-1 branes which are obtained from the regularisation of conical probes. The case of probe ring branes which encircle the background ring branes turns out  not to give an interesting cosmology. All three cases are summarised in Fig.\ref{internalfig}. In this section we will discuss the conditions for slow-roll of the inflaton, how the parameters are compared with observations and a toy mechanism for a graceful exit from inflation. In the remainder of the paper we will not include the dynamics of the modulus field $\psi$, since we have assumed that $T_3 \ll C_i \sim M_*^4$ and therefore it gets stabilised at $\psi_0$ at an energy scale much higher than the scales where inflation takes place and will be of our interest. For this reason, we will set $\psi_0=0$, and restore it whenever we consider it necessary.

\subsection{Slow-roll inflation for the probe brane}

The measure of SUSY breakdown is related to the difference between the brane tensions located
at the conical singularities. As a measure of the deviation from the unwarped case, we may use the parameter
 \be
 \delta \equiv \frac{r^2_0}{r^2_1}-1~,
 \ee
which in terms of the brane tensions, from the brane matching conditions (\ref{tension1}) and (\ref{tension2}), is given by
\be
\d= \frac{T_2-T_1}{2\pi M^4_*-T_2}~. \label{susyb}
\ee

 Let us now consider the conditions under which a probe 3-brane is slowly rolling along the radial direction\footnote{If the radial field is slowly rolling, then one can check that the angular motion is suppressed by the Hubble friction. Of course, there are regions in parameter space where the orbit of the probe codimension-2 brane is spiral, in which case the radial field is fast rolling. As was shown in a similar setup in \cite{angularmotion1} the angular motion is rapidly damped, leading to an almost radial motion of the probe brane. However, it was also pointed out that the perturbations related to the angular motion may lead to an enhancement of isocurvature perturbations\cite{dbihigher}.}. Since we work in an effective four dimensional theory, we can use the standard definitions of the slow-roll parameters
\bea
\epsilon&=&\frac{M^2_P}{2}\bigg(\frac{V'}{V}\bigg)^2~, \\
\eta&=&M^2_P\frac{V^{\prime\prime}}{V}~,
\eea
where $'\equiv d/d\chi$ and $V=V_{\rm probe}$. The above parameters can be found, even without the approximation of a mild warping, as functions of the probe brane position $R$,
\bea
\epsilon&=&\frac{M^2_P}{2}\frac{1}{A^2 W^2}\frac{1}{V^2}\Big(\frac{\partial V}{\partial R}\Big)^2 =\frac{2M^2_P}{r_0^4 T_3}\d^2\frac{R^2}{f^2_0W^{12}}~, \\
\eta&=& M^2_P\frac{1}{A W V}\frac{\partial }{\partial R}\Big(\frac{1}{AW}\frac{\partial V}{\partial R}\Big)
=\frac{2M^2_P}{r_0^2 T_3} \d \frac{1}{W^8}\bigg[1-\d \frac{R^2}{r_0^2 f^2_0W^4}-\frac{2R^2}{r^2_0f_0}\bigg]~,
\eea
where use is made of the relation
\be
\frac{\partial }{\partial R}W^4=\frac{2R}{r_0^2 f^2_0}\d~.
\ee
As can be seen from the above expressions, the slow-roll parameters are proportional to $\d$ and therefore if the warping is small enough,
the brane motion can result to a slow-roll inflation\footnote{When the background is given by the non-conical warped flat solution (\ref{genwarp}) presented in Appendix A, for $|\lambda_3|\ll 1$, the corrections to the slow-roll parameters are given by $\Delta\epsilon={\cal O}( \delta \cdot\lambda_3^2)$ and $\Delta\eta={\cal O}(\delta\cdot\lambda_3) +{\cal O}(\lambda_3^2)$, respectively. Therefore, for $\lambda_3^2\ll |\delta|$, the slow-roll approximation for the inflation remains valid.}.
In terms of the non-canonical field and for any value of $\d$, we have the number of e-foldings as
\be
N= \frac{T_3 }{M^2_P}\int_{R_f}^{R_i} dR \,\frac{A^2W^2 V}{\frac{\partial V}{\partial R}}
=\frac{T_3 r_0^2}{2M^2_P \d}\int_{R_f}^{R_i}\frac{dR}{R}\,W^8~,
\ee
where $R_i$ is the position of the probe at the beginning of inflation and $R_f$ is the position of the probe at the end of inflation.

Let us now see how the above parameters are expressed in terms of the canonical field $\chi$. For $|\delta| \ll 1$, we obtain the canonical scalar field for the radial motion
approximately as
\be
\chi \simeq \chi_0 \arctan \frac{R}{r_0}~,
\ee
with $\chi_0\equiv r_0\sqrt{T_3}$. The field range of $\chi/\chi_0$ is from $0$ to $\frac{\pi}{2}$
for the brane position $R$ running from $0$ to $\infty$.
And in terms of this approximate canonical scalar field, we obtain the probe brane potential as
\be
V(\chi)\simeq T_3\bigg[1+\delta\sin^2\Big(\frac{\chi}{\chi_0}\Big)\bigg]~. \label{potapprox}
\ee
Then we calculate that
\bea
\epsilon&\simeq& \frac{M^2_P\delta^2}{2\chi_0^2}\frac{\sin^2\Big(\frac{2\chi}{\chi_0}\Big)}
{\Big[1+\delta\sin^2\Big(\frac{\chi}{\chi_0}\Big)\Big]^2}~, \label{epsilon}\\
\eta&\simeq&
\frac{2M^2_P\delta}{\chi_0^2}\frac{\cos\Big(\frac{2\chi}{\chi_0}\Big)}{1+\delta\sin^2\Big(\frac{\chi}{\chi_0}\Big)}~.\label{eta}
\eea
Here there are upper bounds on the slow-roll parameters as
\bea
\epsilon&\leq& \frac{M^2_P}{2\chi_0^2}\frac{\delta^2}{1+\delta}\equiv \epsilon_{\rm max}~, \label{simpleeps}\\
|\eta|&\leq& \frac{2M^2_P}{\chi^2_0}|\delta|\equiv |\eta|_{\rm max}~.\label{simpleeta}
\eea
When $\chi^2_0\gg 2M^2_P|\delta|$, the inflaton can be slowly rolling in the entire field space.
In this case, we find the relation between the slow-rolling parameters in most of the field space as
$\epsilon\sim |\delta\eta|$. Thus, for a small $|\delta|$, the $\epsilon$ parameter is negligible compared
to the $|\eta|$ parameter.  We also get the number of e-foldings as
\be
N=\int^{\chi_i}_{\chi_f}\frac{V}{M^2_P V'}d\chi
\simeq - \frac{\chi^2_0}{2M^2_P \delta }\ln \Big(\tan \frac{\chi}{\chi_0}\Big)\Big]\bigg|^{\chi_f}_{\chi_i}~.\label{efold}
\ee
Thus, the number of e-foldings can be sufficiently large, for the field range
of order $\chi_0$.

A similar analysis can be done for the 4-brane that symmetrically encircles the background ring branes. We will not however study in detail the slow-roll parameters for this case because of the following  simple observation. In contrast with the 3-brane slow-roll parameters, the ones for the 4-brane are not proportional to $\d$ and therefore they are not small in general, but only close to special points of the phase space. For the case of the 4-brane with small radius $R_0$ which floats in the internal space, we expect the same kind of dynamics as for the probe 3-brane inflation.

\subsection{Comparison to the observations}

Now we are in a position to discuss about the slow-roll predictions.
The spectral index $n$ is given by the slow-roll parameters as \cite{riottolyth}
\be
n=1-6\epsilon+2\eta~.\label{specind}
\ee
The combined WMAP 5-year data with Baryon Acoustic Oscillations and Type Ia supernovae\cite{wmap5} show that the spectral index is $0.960 \pm 0.013$ ($68\%$ CL). Thus, the spectrum is slightly red-tilted. On the other hand, the adiabatic density perturbations are given  in \cite{riottolyth} by
\be
\delta_H=\frac{1}{\sqrt{75}\pi M^3_P}\frac{V^{3/2}}{V'}=\frac{1}{\sqrt{150}\pi M^2_P}\frac{V^{1/2}}{\epsilon^{1/2}}~,
\ee
where the potential and its derivative are evaluated at the epoch of horizon exit
for the scale $k^{-1}=(a_{\rm COBE}H_{\rm COBE})^{-1}$, which is $H^{-1}_0=3000h^{-1}{\rm Mpc}$, the biggest cosmological scale in the Universe.
From the COBE normalisation we have that $\delta_H=1.91\times 10^{-5}$. Thus, comparing the theoretical prediction from inflation to the observed value, we get a constraint on the inflaton potential as
\be
\frac{V^{1/4}}{\epsilon^{1/4}}=0.027M_P=6.7\times 10^{16} {\rm GeV}~. \label{cobenorm}
\ee

For most of parameter space the slow-roll parameters are $|\eta| \sim |\eta|_{\rm max} = \frac{2M^2_P|\delta|}{\chi^2_0}$ and $\epsilon \sim \epsilon_{\rm max}\sim \frac{1}{4}|\delta|\cdot|\eta|_{\rm max}$.
Then, we see that, in order to have a red tilt, we should have $\delta<0$ and therefore $\eta<0$.
Since the sign of $\eta$ in eq.~(\ref{eta}) changes from negative to positive when crossing the equator at $R=r_0$ or $\frac{\chi}{\chi_0}=\frac{\pi}{4}$, horizon exit should occur before the brane reaches the equator.
So, we get from eq.~(\ref{specind}) the spectral index to be
\be
n= 1-\Big(\frac{3}{2}|\delta|+2\Big)|\eta|_{\rm max}= 1-\Big(\frac{3}{2}|\delta|+2\Big)\frac{\ln(R_f/R_i)}{N}~,
\ee
where use is made of the number of e-foldings obtained from eq.~(\ref{efold}), that is, $N\sim \frac{\ln(R_f/R_i)}{|\eta|_{\rm max}}$.
For $N=(100+75|\delta|)/2\ln(R_f/R_i)$ or $|\eta|_{\rm max}=2/(100+75|\delta|)$, the spectral index is $n= 0.96$.
Thus, $|\eta|_{\rm max}$ should be smaller than $0.02$ to get the spectral index right.
Defining $\alpha\equiv \frac{\chi^2_0}{M^2_P}=\frac{r^2_0 T_3}{M^2_P}$, we have that the condition $T_3<M^4_*$ translates to $\alpha<\frac{1}{\pi\lambda}$, using $M^2_P=\lambda \pi r^2_0 M^4_*$.
Then, since $|\delta|=\frac{\alpha}{2}|\eta|_{\rm max}$, we need to have
\be
|\delta|< \frac{1}{2\pi\lambda}|\eta|_{\rm max}<\frac{0.01}{\pi\lambda}~.\label{warpbound}
\ee
Therefore, we finally  get $|\eta|_{\rm max}\simeq 0.02$ and $N\simeq 50\ln(R_f/R_i)$.
Assuming that inflation starts close to the background ring brane which is at around $R_i=1/M_*$ and taking a typical $R_f=r^2_0 M_*$,
we get the number of e-foldings as $N\simeq 100\ln(r_0M_*)$.
This number of e-foldings is sufficient to account for the
$N_{COBE} \simeq 60$ e-foldings which are necessary to explain the homogeneity of the largest cosmological scale.
Moreover, from eq.~(\ref{cobenorm}), we get the brane tension scale $T_3$ as
\be
T_3^{1/4}=|\delta|^{1/4}\cdot 1.8\times 10^{16} {\rm GeV}\simeq |\delta|^{1/4} M_{\rm GUT}~, \label{cobetension}
\ee
where $M_{\rm GUT}$ is the unification scale in four-dimensional SUSY GUTs.
From the above result and $|\delta|=\frac{\alpha}{2}|\eta|_{\rm max}\simeq 0.01\alpha$, the compactification scale should be
\be
\frac{1}{r_0}= 0.1\frac{M^2_{\rm GUT}}{M_P}\simeq 10^{13}{\rm GeV}~.\label{compactscale}
\ee
Therefore, the COBE normalization and the spectral index require a rather low compactification scale.
 However, the precise value of $\delta$ or the probe brane tension scale from $|\delta| \simeq 0.01\alpha$ is not determined.

Another observable is the running of the spectral index. According to \cite{riottolyth}, it is given by
\be
{dn \over d \ln k}= 24 \epsilon^2 -16 \epsilon \eta +2 \xi^2~,
\ee
where the new parameter is
\be
\xi^2=M_P^4 {V' V''' \over V^2}~.
\ee
Evaluating the above quantity for (\ref{potapprox}), we obtain
\bea
{dn \over d \ln k} &\simeq& {M_P^4 \over \chi_0^4} \d^2 \bigg[  6 \d^2 {\sin^4(2 \chi/\chi_0) \over (1+ \delta \sin^2 (\chi/\chi_0))^4} -16  \d {\sin^2(2 \chi/\chi_0) \cos(2 \chi/\chi_0) \over (1+ \delta \sin^2 (\chi/\chi_0))^3} \nonumber
\\
&&\quad\quad-8  {\sin^2(2 \chi/\chi_0) \over (1+ \delta \sin^2 (\chi/\chi_0))^2}   \bigg]~.
\eea

The last term in the above expression comes from $\xi^2$ and it dominates for small $\d$. As order of magnitude, in most of the field space (\ie away from special points), we have  $dn/ (d \ln k) \sim \eta^2 \sim 10^{-4}$.  Thus, the observational constraint \cite{wmap5} is easily satisfied.

The final observable we are going to mention is the ratio of the tensor to scalar perturbations. It is given by $r=12.4 \epsilon$ and thus it is $r \simeq 3.1|\delta\cdot \eta|\simeq 0.062|\delta|$. Therefore, for $|\delta|<\frac{0.01}{\pi\lambda}$, we get $r<0.0062/(\pi\lambda)$ which is in agreement with observations \cite{wmap5}, $r<0.22~ (95\% {\rm CL})$ for power-law models.

\subsection{Graceful exit from inflation}

As mentioned in the beginning of this section, we have assumed that inflation ends when the probe brane hits the background brane at the other end. The exact dynamics at the collision point is beyond the limitations of this model.
However, in order for the  model to be realistic, we know that there should be on the one hand a mechanism of reducing the vacuum energy to zero and on the other hand a mechanism which reheats the brane. Here, we will consider a toy model featuring both mechanisms which need to occur at the end of inflation, thus providing a way of having a so called graceful exit. We regard this toy model as a crude approximation of the physics taking place at the brane collision.

For this purpose we will consider a possibility of having an analogue of hybrid inflation \cite{hybrid} in our probe codimension-2 brane scenario. To this, we add a real scalar field $\varphi$ with a coupling to the probe brane. For instance, we take the effective potential for the system with the inflaton and the real scalar field (the waterfall field) to be
\be
V_{\rm eff}= (T_3 + \alpha^2\varphi^2) W^4(R) + h (\varphi^2-\beta^2)^2-h \beta^4 \label{hybrid}
\ee
where we assumed that the $\psi$ modulus has been stabilised before inflation starts.
When the waterfall field lives on the probe brane, one should choose a specific dilaton coupling for obtaining the self couplings in the latter part of eq.~(\ref{hybrid}). On the other hand, when the waterfall field lives in the bulk, the self couplings with arbitrary dilaton dependence can be introduced on the background brane.
Then, in either cases, as long as the brane position is such that  $\alpha^2 W^4(R)>2h\beta^2$, the waterfall field gets a vanishing VEV, so the inflation as discussed previously takes place. We need that the above condition is satisfied for the motion of the probe brane until it comes close to the point of collision with the background ring brane. Close to that point, the direction of the inequality will be reversed $\alpha^2 W^4(R)<2h\beta^2$ and the waterfall field starts rolling down to a minimum with nonzero VEV, thus ending  inflation. For the above scenario to work, we should carefully choose the parameters $h$, $\a$ and $\b$ so that the
 waterfall field starts fast-rolling to its new minimum (and reheats the brane) after the   $N>60$ e-folds of inflation. Furthermore,  in this case, we can fine-tune the parameter $\beta$, such that the vacuum energy at the minimum vanishes.

\section{The backreaction of the probe brane}

In the previous section, by analysing the slow-roll conditions combined with the COBE normalization we could not determine the
precise value of the parameter $\delta$, but only the upper bound on $|\delta|$.
When the value of $\alpha=\frac{r^2_0 T_3}{M^2_P}$ is fixed, the required value of $|\delta|$ is also determined
by $|\delta|\simeq 0.01\alpha$. When $\alpha$ is of order unity, \ie $T_3\sim 10^{-2} M^4_{GUT}$, we only need a small amount of tuning to get $|\delta|\simeq 0.01$. However, the probe brane with such a large tension could give a large backreaction, such that the background with static extra dimensions and a mild warping could be deformed significantly. In this section, we discuss the backreaction issues and find the conditions required for justifying the probe brane approximation.

\subsection{Backreaction on the volume modulus}

In our previous analysis, we have first assumed that the backreaction of the probe brane is negligible for the modulus stabilization. This would be the case if the modulus is stabilised at a scale higher than the scale of inflation. This is equivalent with demanding that the Hubble scale during the inflation is smaller than the modulus mass  $H^2 \ll m^2_{\psi}$,
where $H^2=\frac{T_{3}}{3M^2_P}e^{-2\psi_0}$, with $m^2_\psi$ being the modulus mass
given in eq.~(\ref{modulusmass}) [in this subsection, we  momentarily restore $\psi_0$ in our equations]. Thus, the probe brane approximation is justified under the condition,
\be
T_3\ll 3 e^{2\psi_0}M^2_P m^2_\psi~.\label{backreaction}
\ee
For instance, when the dilaton coupling parameters of the regularized background branes are given by $(s_2,t_2)=(3,0)$ and $(s_1,t_1)=(\frac{1}{2}(1-\varepsilon),3)$ with $\varepsilon\ll 1$,
using eqs.~(\ref{egmodulusvev}) and (\ref{egmodulusmass}), the condition (\ref{backreaction}) becomes
\be
T_3\ll \frac{27}{8}\Big(\frac{2C_2}{9C_1}\Big)^{1/5} C_1~.
\ee
For $C_1\sim C_2\sim M^4_*$ we get $T_3\ll M^4_*$.
Thus, the constraint on the warping $|\delta|\simeq 0.01\alpha$ in (\ref{warpbound}) becomes more severe as follows,
\be
|\d| \ll \frac{0.01}{\pi \la}~.
\ee

A natural way to make the probe brane tension much smaller than  the six-dimensional Planck scale is to consider it slightly non-conical. To see that, let us consider the probe brane as being a ring 4-brane with a small radius and with a small violation of the scaling symmetry, so that the limiting 3-brane is non-conical. Since the ring brane has a non-conical thin limit, we expect that its presence in the bulk will backreact to the four dimensional part of the metric so that the latter obtains an expansion rate of the order of $H^2=\frac{T_{3}}{3M^2_P}$. This is to be contrasted with the property of conical branes, which are able to "hide" vacuum energy in a local deficit angle without curving their world-volume \cite{CLP}.

To obtain the correction to the potential for the inflaton in this new setup, we should study a six-dimensional system  with curved (de-Sitter) four-dimensional foliations. We refer the interested reader to Appendix C for an extended analysis of this case. Here, let us sketch briefly how the derivation of the final result is obtained. In Ref.~\cite{curved}, solutions of the six-dimensional system with curved (de-Sitter) four-dimensional foliations and axial symmetry were discussed. The setup in \cite{curved} had two branes, at least one of which was non-conical. In Appendix C, we extend the analysis of \cite{curved} for the model that we are considering, \ie two background branes at the axis of symmetry and one non-conical probe brane. The curved solutions around the probe brane  can be expanded around a conical flat solution with warp factor $W_f$ and dilaton $\phi_f=4\ln W_f$. At this point,  we consider the inflation potential from an effective action approach (with the modulus stabilised at some high scale), as we did for the modulus potential in Sec. 3, and take couplings for the ring-regularised non-conical brane which marginally break the scale invariance, as for example  $V_p=v_p e^{-\frac{1}{4}s\phi}$ with $|1-s|\ll 1$. After we impose a condition on the asymptotic near-brane solution, the effective potential reduces to
\be
V_{\rm eff}\simeq T_3 W^4_f \Big[1+(1-s)\ln W_f\Big] \label{nonconeff}
\ee
where $T_3\equiv \frac{\pi}{2}(1-s)v_p R_0$ is the effective probe brane tension for the small $R_0$ limit. As a result, for a small breaking of scaling invariance, the effective 3-brane tension is naturally suppressed compared to the six dimensional  fundamental scale\footnote{We take both the 4-brane tension and the 4-brane radius to be of order the six dimensional fundamental scale.}. Therefore, in the presence of a steep modulus potential coming from the large breaking of scale invariance of the background branes, it is possible to ignore the backreaction of the probe brane tension on the volume modulus.

Moreover, the correction term to the DBI action becomes negligible. The inflaton potential that we obtain in this effective action approach, is the same (modulo terms having to do with the deviation form the conical limit) as the one obtained from the DBI action for the probe brane with scaling invariance. Therefore, the DBI action approximation, used everywhere in the previous sections, holds. Furthermore, the above solves an apparent paradox that the careful reader may have noticed. As is known, codimension-2 branes can have peculiar properties regarding the relation between their energy content and their curvature. We mentioned above that conical branes can "hide" vacuum energy in a local deficit angle without  curving themselves. In such a case, the potential energy needed for inflation will not gravitate as expected from the DBI action. This paradox does not hold anymore if we depart from the conical limit, since then the potential energy of the probe will necessarily curve its worldvolume.

\subsection{Backreaction on the warp factor}

In the previous subsection, we have seen that the effective potential for a probe (slightly non-conical) brane, with an expansion rate much lower than the compactification scale, can be in a good approximation given by the one obtained from the flat brane warp factor. The mild warping that we need for a successful inflationary period, can be realised if the background brane tensions are almost equal, as we can see from eq.~(\ref{susyb}). In order that the probe brane approximation holds, we need to verify that the probe brane tension does not  change much the warp factor generated by the imbalance between the two background branes. An easy way to make sure that this is not happening is to demand that the probe brane tension is much smaller than the difference between the background brane tensions, \ie $T_3 \ll |T_2-T_1|$. This is equivalent to imposing a condition on the deficit angle of the background branes as $\la \pi \ll 10^{-2}$. In this case, the background branes have tension close to critical, \ie the deficit angle is close to $2\pi$. However, a tuning of $\la$ like this, is an extra tuning besides the one of $\d$, which we would like to avoid.

In the following, we will concentrate on the possibility that $\la \sim {\cal O}(1)$ and see if a probe brane tension larger than the difference  between the background brane tensions, \ie $T_3 > |T_2-T_1|$, can still give a small backreaction to the warp factor (and therefore the inflationary potential). In this section, we discuss this backreaction by looking at how the warp factor reacts to the additional brane(s) with an explicit multiple brane solution.

Multi-brane solutions for the supergravity system that we study are analysed in \cite{multibranes}. There, it is shown that
it is possible to have general warped flat solutions with multiple conical branes, if an undetermined holomorphic function $V(z)$ appearing in  the metric solution has multiple zeros \cite{multibranes}. Thus, we can add
more branes in our warped background and identify one of the additional branes as the probe brane that was used for driving inflation. In Appendix D we summarise the multiple brane solution with more than two background branes and we link it to the background model that we used earlier on.

Here let us briefly mention the solution without going into many details. The four-dimensional warp factor in the general warped solution with multiple branes can be written in the following form \cite{multibranes}
\be
ds^2= \hat{W}^2 \eta_{\m\n} dx^\m dx^\n ++\frac{1}{2|V(z)|^2}\frac{P({\hat W})}{{\hat W}^2}dz d{\bar z}~,
\ee
with $\hat{W}$ the multibrane solution warp factor, $P(\hat{W})$ a definite function of the warp factor given in Appendix D and $V(z)$  an arbitrary holomorphic function of $z$. By choosing different holomorphic functions $V(z)$, we can generate solutions where the branes vary in numbers and in characteristics, as their tension and position in the bulk. For small warping, the additional branes will not change the warp factor much, but instead they will induce additional local deficit angles around them. By choosing this holomorphic function as
\be
V(z)=\frac{z}{|c|}\left(1+{i\beta \over z+z^{-1} }\right)~, \label{holo}
\ee
we can construct a system of branes which generates a warp factor similar to the one we discussed in the previous text. In particular, the holomorphic function (\ref{holo}) generates a system of six branes: two branes have fixed tensions $T_{\pm i}=-2\pi M^4_*$ at $z=\pm i$, two branes with tensions given by (\ref{tension1}), (\ref{tension2})
\be
T_1=2\pi M^4_*(1-\lambda)~, \quad T_2=2\pi M^4_*\Big(1-\lambda\frac{r^2_1}{r^2_0}\Big)~,
\ee
at $z=0,\infty$ and two additional branes with varying tension given by
\be
T_3=2\pi M^4_*(1-|b| \lambda)~, \quad T_4 = 2\pi M^4_* \Big(1-|b|\lambda \frac{r^2_1}{r^2_0}\Big)~, \label{newbranes}
\ee
at $z_\pm = -{i \over 2}(\b\pm \sqrt{\b^2+4})$. The parameters in the brane tensions are related to the parameters of the holomorphic function as, $\la={1 \over 4}|c|g^2({r_0^2 \over r_1^2}-1)$ and $|b|={|\b| \over \sqrt{\b^2 +4}}$. Then, from the above, we get a relation between the additional brane tensions and the warping,
\be
\frac{2\pi M^4_*-T_3}{2\pi M^4_*-T_4}=\frac{r^2_1}{r^2_0}~.\label{addtune}
\ee
For the  mild warping $\frac{r_1}{r_0}\sim 1$ needed for the slow-roll inflation, eq.~(\ref{addtune}) requires that a would-be probe brane with tension $T_3$ should be paired with an additional brane with almost the same tension as $T_4\sim T_3$.
However, since the brane tensions (\ref{newbranes}) depend on an arbitrary local deficit angle parameter $|b|$, we can dissociate the tensions $T_{1,2}$ from $T_{3,4}$.

We may regard the multiple brane solution as an adiabatic approximation of the moving probe branes.
Suppose that the would-be probe branes with positive tension start rolling at some bulk positions. Then, the probe brane inflation ends when the probe branes hit the would-be background branes at $z=0$ and $|z|=\infty$, respectively, for $|b|=1$ or $|\beta|=\infty$. Then, the background solution settles into another flat solution with two brane tensions only, $T'_1=T_1+T_3=2\pi M_*^4(1-\lambda')$ and $T'_2=T_2+T_4=2\pi M_*^4 \Big(1-\lambda'\frac{r^2_1}{r^2_0}\Big)$ by changing the deficit angle parameter from $\lambda$ to $\lambda'=2\lambda-1$.
Thus, after the probe branes hit the background branes, it would be natural to get a graceful exit from inflation.

Finally, let us consider the flux quantization condition for multi-brane solutions to see if the tuning required for a mild warping may be affected by the additional branes. Since the flux quantization condition relates the probe brane tension with the background brane tensions, the effect of the probe on the relation between the background brane tensions (and thus the warp factor) is expected to be generically strong. However, as shown in Appendix D, if we assume that one of the background branes is non-BPS (due to the VEV $\langle Q_1\rangle$ of a brane scalar field with R charge $r_1$),
the flux quantisation condition reads
\be
\Big(2-\frac{T_1+T_3}{2\pi M_*^4}\Big)\Big(2-\frac{T_2+T_4}{2\pi M_*^4}\Big)
= \bigg[n-\frac{1}{4\pi M_*^4}\Big(T_1+2r_1g^2 M_*^4|\langle Q_1\rangle|^2+ T_2\Big)\bigg]^2~.
\ee
In the above equation, we would like to keep the  background brane tensions $T_{1,2}$ fixed, in order to ensure that the four-dimensional warp factor does not change with the introduction of the extra bulk branes. Then, by dynamically tuning the brane scalar VEV $\langle Q_1\rangle$, we can tune the effective localized FI term of the background brane, such that the probe brane tension does not change the relation between the background brane tensions needed for a mild warping. This tuning is independent of the tuning needed for ensuring slow-roll.

In conclusion, we see from the above, that one can maintain the small tuning between the background branes needed for the mild warping by using corrections to its FI term due to the VEV of a brane scalar field. This shows that even though the probe brane tension could be larger than the difference  between the background brane tensions, we can still have a small backreaction to the warp factor.

\section{Conclusion}

We have pursued the possibility of realizing inflation from a probe non-BPS codimension-2 brane moving in a warped background
in the context of six-dimensional gauged supergravity. The probe codimension-2 brane gets a nonzero potential
due to the warping of the four-dimensional metric. Unlike the probe D-brane case \cite{KKLT}, the slow-roll inflation is possible for a mild warping. In turn, from the brane matching conditions, it is necessary to have two background branes with almost equal tensions for a mild warping.
Then, the nonzero inflaton potential can be understood due to the attractive interaction between the probe brane with lower tension and a background brane. For the compactification scale of order $10^{13}$ GeV, we can match both the COBE normalization and the spectral index as observed by WMAP.

The key assumptions that we made for a successful probe-brane inflation are listed the following:
\begin{itemize}
\item The background brane tensions are tunable, such that the brane tension difference is made small enough, being close to a SUSY vacuum;
\item Upon regularising the background branes as ring-like branes, the scale invariance of the equations of motion is {\it badly} broken on the background branes, due to the brane dilaton potentials. This stabilises the volume modulus at a high energy scale without affecting the background geometry;
\item The bulk hyperscalars and the  Kalb-Ramond field get large masses on the background branes or they do not couple to the probe brane;
\item The probe brane is {\it slightly} non-conical, achieved with a {\it slight} breaking of the scale invariance on the probe brane, for suppressing its tension in comparison to the six-dimensional fundamental scale and also for  making the DBI action a good approximation;
\item Graceful exit occurs at the brane collision and is described by the coupling of a tachyonic scalar field to the probe brane.
\end{itemize}

Considering the backreaction of the probe brane on the volume modulus, we found that the probe brane tension should be much smaller than the six-dimensional fundamental scale. As stated above, this can be achieved with considering a probe brane which is slightly non-conical. For this  supergravity model,  there exist warped brane solutions with four-dimensional  de-Sitter space \cite{curved}, for which at least one of the branes must be non-conical. We  computed the four-dimensional effective action for a non-conical probe brane and showed that it reproduces the DBI action for a conical 3-brane. Furthermore, it is straightforward to see that this effective action is consistent with the presence of the six-dimensional warped de-Sitter brane solution. The required suppression of the probe brane tension (with respect to the six-dimensional Planck mass) determines the amount of tuning needed for the background brane tensions for obtaining a small warping as $|\delta|\ll \frac{0.01}{\pi\lambda}$. Moreover, from an explicit multiple brane solution, we showed that the backreaction of an additional brane on the warp factor can be fully taken into account, with no modification of the tuning needed between the background branes in order to have small warping. All the above arguments confirm that the probe brane approximation used in this brane inflation scenario is valid.

Concluding, the codimension-two probe brane inflation scenario could be realized in a controllable way, provided that the background branes are tuned between each other for a mild warping. A lot of work has to be still done regarding phenomenology in order to test it against observations. An interesting direction of future research would be to compute the non-Gaussianities  of the model, as this is known from DBI inflation models (see for example \cite{dbihigher}) to provide testable predictions. What we can surely say by the present work, is that due to the negligible amount of gravitational waves predicted by the model, it is in principle falsifiable if a significant amount of them is observed by near-future observations.

\section*{Acknowledgments}
We would like to thank Cliff Burgess, Kazuya Koyama, S\'ebastien Renaux-Petel, Subramaniam Shankaranarayanan and Andrew Tolley  for discussions.
H.M.L. is supported by the research fund from the Natural Sciences and Engineering Research Council (NSERC) of Canada. A.P. is supported
by a Marie Curie Intra-European Fellowship EIF-039189.

\def\theequation{A.\arabic{equation}}
\setcounter{equation}{0}
\vskip0.8cm
\noindent
{\Large \bf Appendix A: Non-conical warped background}
\vskip0.4cm
\noindent

The general warped solution with four-dimensional Minkowski space and internal axial symmetry is generalized to the singular one with non-conical branes as following \cite{gibbons},
\bea
ds^2&=&W^2(\eta)\eta_{\mu\nu}dx^\mu dx^\nu +K^2(\eta)W^8(\eta)d\eta^2+K^2(\eta)d\theta^2~, \nonumber \\
\phi&=&4\ln W +2\lambda_3\eta~, \nonumber \\
F_{\eta\theta}&=&q e^{-\frac{1}{2}\phi}W^{-4}\epsilon_{\eta\theta}~, \label{genwarp}
\eea
where
\bea
W^4&=&\bigg|\frac{q\lambda_2}{4g\lambda_1}\bigg|\,\frac{\cosh[\lambda_1(\eta-\xi_1)]}{\cosh[\lambda_2(\eta-\xi_2)]}~, \\
K^{-4}&=&\bigg|\frac{gq^3}{\lambda_1^3\lambda_2}\bigg|e^{-2\lambda_3\eta}\cosh^3[\lambda_1(\eta-\xi_1)]\cosh[\lambda_2(\eta-\xi_2)]~.
\eea
Here, we note that $\lambda^2_2=\lambda^2_1+\lambda^2_3$.

By making a coordinate transformation $r=e^{\lambda_1\eta}$, with $r_0=e^{\lambda_1\xi_1}$ and $r_1=e^{\lambda_1\xi_2}$,
we can rewrite the singular warped solution in the new coordinate as
\bea
ds^2&=&W^2(r)\eta_{\mu\nu}dx^\mu dx^\nu +A^2(r)(dr^2+B^2(r)d\theta^2)~, \\
\phi&=&4\ln W+\frac{2\lambda_3}{\lambda_1}\ln r~, \\
F_{r\theta}&=&q e^{-\frac{1}{2}\phi}W^{-4}\epsilon_{r\theta}~,
\eea
where
\bea
W^4&=&\Big(\frac{r}{r_0}\Big)^{\frac{\lambda_2}{\lambda_1}-1}\frac{f_1}{{\tilde f}_0}~, \\
A&=&r^{\frac{\lambda_3}{2\lambda_1}}\Big(\frac{r}{r_0}\Big)^{\frac{\lambda_2}{\lambda_1}-1}\frac{W}{{\tilde f}_0}~,
\quad B=\frac{\lambda_1r}{W^4}~,\\
{\tilde f}_0&=&1+\Big(\frac{r}{r_0}\Big)^{\frac{2\lambda_2}{\lambda_1}}~,\quad
f_1=1+\frac{r^2}{r^2_1}~.
\eea
The two radii $r_0,r_1$ are given by
\be
r^2_0=\frac{\lambda^2_2}{\lambda^2_1g^2 M_*^4}~, \quad r^2_1=\frac{4 M_*^4}{q^2}~.
\ee
Here we have rescaled the four-dimensional and radial coordinates as well as two radii such that $W$ and $A$ get the forms as above.
We note that for $\lambda_3=0$, \ie $\lambda_1=\lambda_2$, we reproduce the warped solution with two conical branes.

In the above we have set the arbitrary integration constant of $\f$ to zero. This can be restored by a simple scale transformation $g_{MN}\rightarrow e^{\frac{1}{2}\phi_0}g_{MN}$ and $\phi\rightarrow \phi+\phi_0$.

\def\theequation{B.\arabic{equation}}
\setcounter{equation}{0}
\vskip0.8cm
\noindent
{\Large \bf Appendix B: The effect of a nonzero localized Fayet-Iliopoulos term}
\vskip0.4cm
\noindent

In this Appendix,
we discuss the effect of a localized FI term $\xi_3$ corresponding to a moving probe brane with nonzero tension $T_3$.
We show that an electric source is needed for satisfying both the equation of the bulk gauge field and the Bianchi identity.
 Then, the localized FI term, combined with the electric source, gives a divergent contribution to the inflaton kinetic term.

In the presence of the probe brane, the gauge field strength is modified to
\be
{\hat F}_{mn}=\langle {\hat F}_{mn}\rangle-\Big(\frac{2}{M^4_*}\Big)^{1/2}\xi_3\epsilon_{mn}\frac{\delta^2(y-y_p(t))}{e_2}~.
\ee
Then, if $A_\theta$ is the only nonzero component of the gauge field,
the relevant equation for the bulk gauge field becomes
\be
\partial_r(\sqrt{-g}e^{\frac{1}{2}\phi}{\hat F}^{r\theta})+\partial_0(\sqrt{-g}e^{\frac{1}{2}\phi}F^{0\theta})
=-\frac{\partial {\cal L}_{e}}{\partial A_\theta}~,\label{gaugetheta}
\ee
and
\be
\partial_\theta(\sqrt{-g}e^{\frac{1}{2}\phi}F^{\theta 0})=-\frac{\partial {\cal L}_e}{\partial A_0}~,
\ee
where ${\cal L}_e$ is the Lagrangian for an electric source.
On the other hand, the Bianchi identity for the bulk gauge field reads
\be
\partial_\theta F_{r\theta}+\partial_r F_{\theta 0}=0~.\label{bi}
\ee
We will take the solution for $F_{r\theta}$ as
\be
F_{r\theta}=\langle F_{r\theta}\rangle +\Big(\frac{2}{M^4_*}\Big)^{1/2} \xi_3\epsilon_{r\theta}\frac{\delta^2(y-y_p(t))}{e_2}~,
\ee
where $\langle F_{r\theta}\rangle$ is the background flux.
The last term of the above equation, for $y_p=(R(t),\theta_0)$, is written as
\be
\epsilon_{r\theta}\frac{\delta^2(y-y_p(t))}{e_2}= \delta(r-R(t))\delta(\theta-\theta_0)~.
\ee
Then, the angular component of the gauge equation (\ref{gaugetheta}) becomes
\be
\partial_0(\sqrt{-g}e^{\frac{1}{2}\phi}F^{0\theta})=-\frac{\partial{\cal L}_e}{\partial A_\theta}~.
\ee
Therefore, we choose the action for the electric source to be $S_e=\int d^6x {\cal L}_e$ with
\be
{\cal L}_e= A_\theta J^\theta+ A_0 J^0~,
\ee
where $J^\theta=-\partial_0(\sqrt{-g}e^{\frac{1}{2}\phi}F^{0\theta})$
and $J^0=\partial_\theta (\sqrt{-g}e^{\frac{1}{2}\phi}F^{0\theta})$.
 The solution for $F_{0\theta}$ is obtained from solving the Bianchi identity (\ref{bi}) as
\be
F_{0\theta}=-\Big(\frac{2}{M^4_*}\Big)^{1/2} \xi_3 {\dot R}\delta(r-R(t))\delta(\theta-\theta_0)~.
\ee
That is, for $A_0=0$, the gauge potential satisfying $A_\theta=\langle A_\theta\rangle$ for $r<R(t)$ is obtained as
\be
A_\theta = \langle A_\theta\rangle + \Big(\frac{2}{M^4_*}\Big)^{1/2} \xi_3\theta(r-R(t))\delta(\theta-\theta_0)~.
\ee

Let us now consider the contribution of the FI term and the electric source to the effective action.
Due to the time-dependent piece in the gauge potential, the additional term in the effective action is
\bea
\Delta S_{\rm eff}&=&\int d^6 x \sqrt{-g}\bigg[-\frac{1}{2}e^{\frac{1}{2}\phi}F_{0\theta}F^{0\theta}
+\frac{1}{\sqrt{-g}}(A_\theta J^\theta+A_0 J^0)\bigg] \nonumber \\
&=&\int d^6 x \sqrt{-g}\bigg[-\frac{1}{2}e^{\frac{1}{2}\phi}F_{0\theta}F^{0\theta}
-\frac{1}{\sqrt{-g}}A_\theta\partial_0(\sqrt{-g}e^{\frac{1}{2}\phi}F^{0\theta})\bigg]~.
\eea
After integration by parts, the above additional term becomes
\bea
\Delta S_{\rm eff}&=&\int d^6 x \sqrt{-g}\,\frac{1}{2}e^{\frac{1}{2}\phi}F_{0\theta}F^{0\theta} \nonumber \\
&=&-\int d^4x \sqrt{-{\tilde g}} \, \frac{1}{M^4_*} \xi^2_3\,e^{\frac{1}{2}\phi_0}\frac{W^8(R)}{\lambda R} {\dot R}^2
\delta(r=0)\delta(\theta=0)~.
\eea
Therefore, the FI term of the probe brane would give rise to a divergent correction to the kinetic term of the inflaton. This can be regularised by thickening the brane. After taking into account this regularization, and for a mild warping, the FI term correction becomes roughly $-\k \frac{2}{M^4_*}\xi^2_3\frac{{\dot R^2}}{R}$, with unknown coefficient $\k$ so the total kinetic term would be $K=(\frac{T_3}{f^2_0(R)}-\frac{\k\xi^2_3}{R} \frac{2}{M^4_*}){\dot R}^2$. Thus, in order for the inflaton not to develop any ghost instability for $\k>0$, the inflaton must lie in the range satisfying $\frac{\k\xi^2_3}{r_0 T_3} \frac{2}{M^4_*} <\frac{\frac{R}{r_0}}{1+\frac{R^2}{r^2_0}}$.
So, for $\frac{\k\xi^2_3}{r_0 T_3} \frac{2}{M^4_*} \ll 1$, we can still get a proper field range of order $\phi_0$ for the inflaton such that the discussion in section 5 still holds.

\def\theequation{C.\arabic{equation}}
\setcounter{equation}{0}

\vskip0.8cm
\noindent
{\Large \bf Appendix C: Effective action approach for the probe brane potential}
\vskip0.4cm
\noindent

In this Appendix, we will compute the inflaton potential for a non-conical probe brane from an effective action approach. This will serve two purposes. First, we can see that for a small departure from "conicality" the probe brane tension is naturally suppressed compared to the six dimensional fundamental scale. Secondly, we can confirm the validity of the DBI action when discussing the effective four-dimensional theory. We first generalize the analysis  of four-dimensional curved solution with two branes of \cite{curved}, to the case with multiple branes (since in our setup we have two background branes and a non-conical probe brane). Here, we assume that the local behavior of the curved multiple brane solution at the brane is similar to the one of the curved solution with two branes \cite{curved}, as it is true of the flat solution \cite{multibranes}.

In a coordinate patch adapted to a brane $(r_i,\o_i)$, the metric will be
\be
ds^2 = W^2(r_i,\o_i) g_{\m\n}(x) dx^\m dx^\n + A^2_i(r_i,\o_i)(dr_i^2 +B^2_i(r_i,\o_i) d\o_i^2)
\ee
but, in the close neighborhood of the brane, the metric will be approximately axially symmetric, \ie the above functions will not have strong dependence on $\o_i$ (however, as we saw in Sec. 4.2  this small dependence may be crucial for obtaining the right potential).
Then, we see that the six-dimensional Einstein and dilaton equations imply a general relation between the warping at the branes and the four-dimensional curvature\cite{curved} as
\be
\sum_i \la_{1i} r_i \frac{\partial}{\partial r_i}\Big(\ln W-\frac{1}{4}\phi\Big)\Big|_{r_{i,b}}=\frac{1}{8\pi}R_4 V_2. \label{curvedcond}
\ee
Here $r_{i,b}$ is the brane position, $R_4=12H^2$ is the curvature for a four-dimensional  de-Sitter metric $g_{\mu\nu}(x)$ and $V_2=\int d^2y e_2 W^2$ is the volume of extra dimensions. For a small Hubble scale compared to the compactification scale,
we take the flat conical solution to be the dominant piece of the curved solution at the ring brane positions $r_i = r_{i,b}$
and add the correction terms coming from the asymptotic limits of the curved solution \cite{curved} as
\bea
\ln W &\rightarrow& \ln W_f+\frac{1}{4}\Big(-1+\frac{\lambda_{2i}}{\lambda_{1i}}\Big)\ln r_i , \nonumber \\
\ln (A_iB_i) &\rightarrow& \ln(A_f B_f)+\frac{1}{4}\Big(-1+\frac{\lambda_{2i}}{\lambda_{1i}}+\frac{2\lambda_{3i}}{\lambda_{1i}}\Big)\ln r_i, \label{expansion}\\
\phi &\rightarrow& \phi_f+\Big(-1+\frac{\lambda_{2i}}{\lambda_{1i}}+\frac{2\lambda_{3i}}{\lambda_{1i}}\Big)\ln r_i \nonumber
\eea
where $\lambda_{1i},\lambda_{2i},\lambda_{3i}$ are constant parameters and they satisfy $\lambda^2_{2i}=\lambda^2_{1i}+\lambda^2_{3i}$, $\lambda_{2i}>0$ and $2\lambda_{2i}+\lambda_{3i}>0$.
Here $W_f,A_f,B_f,\phi_f$ are the functions obtained for the conical flat solution with multiple branes\cite{multibranes}.
When two background branes have conical limits, $\lambda_{3i}=0$ and $\lambda_{1i}=\lambda_{2i}$ for $(i=1,2)$, so that $\phi\rightarrow 4\ln W_f$ at those branes. In this case, the junction conditions at the background branes will give no contribution on the left-hand side of eq.~(\ref{curvedcond}). Then, for a nonzero four-dimensional curvature, we would need $\lambda_{3p}\neq 0$ for the probe brane. Then, similarly to the junction conditions for the regularized background branes in eq.~(\ref{junctions}), for the probe brane action with dilaton couplings $V_p,U_p$, the junction condition at the probe brane becomes
\be
\lambda_{3p} = W^4 A_p B_p \bigg[\frac{V_p}{2}+2\frac{\partial V_p}{\partial\phi}
-\frac{k^2_p}{2A_p^2B_p^2}\Big(\frac{U_p}{2}-2\frac{\partial U_p}{\partial\phi}\Big)\bigg]\bigg|_{r_{p,b}} \label{lambda3}
\ee
where we set the $U(1)_R$ gauge coupling to the probe brane to zero for simplicity.

In the presence of the probe brane, we can derive the four-dimensional effective potential, as we did in Sec. 3 for finding the effective potential for the modulus $\psi$. In the present case, the effective action will depend, apart from the modulus, on a new modulus which is the position of the probe brane $R$, \ie the inflaton. For the inflation and compactification scales that we have taken before, the modulus is fixed at a scale much higher than the inflation scale. Therefore, fixing $\psi$ the effective potential will be a function of the probe brane position $R$.
The effective potential for the probe brane is
\be
V_{\rm eff}=\pi \int d\omega_p W^4 A_p B_p\bigg[\frac{V_p}{2}+2\frac{\partial V_p}{\partial\phi}
-\frac{k^2_p}{2A_p^2B_p^2}\Big(\frac{U_p}{2}-2\frac{\partial U_p}{\partial\phi}\Big)\bigg]\bigg|_{r_{p,b}}.
\ee
Here, using the expansions of the solution around the probe brane from eq.~(\ref{expansion}), we obtain
\be
W^4A_p B_p e^{-s\f/4} \Big|_{r_{p,b}}=W^4_f A_f B_f e^{-s\f_f/4}
r_p^{-1+\frac{\lambda_{2p}}{\lambda_{1p}}+\frac{1}{4}(1-s)(-1+\frac{\lambda_{2p}}{\lambda_{1p}}+\frac{2\lambda_{3p}}{\lambda_{1p}})}.
\ee
In Section 4.2, we argued that $\int d\o_p A_f B_f  e^{-\f_f/4}  W^4_f \approx 2 \pi R_0  W^4_f$. Therefore, choosing $\lambda_{2p}=\lambda_{1p}-\frac{1-s}{5-s}\cdot 2\lambda_{3p}$,
we can rewrite the four-dimensional effective potential as
\be
V_{\rm eff}=\pi R_0 W^4_f e^{(1-s)\f_f/4}e^{s\f/4} \bigg[\frac{V_p}{2}+2\frac{\partial V_p}{\partial\phi}
-\frac{k^2_p}{2A_p^2B_p^2}\Big(\frac{U_p}{2}-2\frac{\partial U_p}{\partial\phi}\Big)\bigg]\bigg|_{r_{p,b}}
\ee
where we have implicitly taken into account the delicate $\o_p$ integration. In this case, with the condition $\lambda^2_{2p}=\lambda^2_{1p}+\lambda^2_{3p}$, we would require
\be
\la_{1p}= -{1 \over 4} {(s+3)(7-3s) \over (1-s)(5-s)}\la_{3p} \ \ , \  \  \la_{2p}=-{8 (s-1)^2 +(s+3)(7-3s) \over 4 (1-s)(5-s)}\la_{3p} ~.
\ee

To see how the mechanism can work, let us take a simple example with $V_p=v_p e^{-\frac{1}{4}s\phi}$ and $U_p=u_p e^{\frac{1}{4}\phi}$. Then, the
 above effective potential becomes
\be
V_{\rm eff}=T_3 W^4_f e^{\frac{1}{4}(1-s)\phi_f}\bigg|_{r_{p,b}}\label{nonconeffa}
\ee
where $T_3\equiv \frac{\pi}{2}(1-s)v_p R_0$ is the effective probe brane tension for the small $R_0$ limit. Moreover, from eqs.~(\ref{curvedcond}) and (\ref{lambda3}), we obtain
\be
\pi\lambda_{3p}=\frac{1}{2}\pi W^4 A_p B_p (1-s)v_p e^{-\frac{1}{4}s\phi}\bigg|_{r_{p,b}}=3H^2 V_2.
\ee
For $H^2\simeq\frac{T_3}{3M^2_P}$ and $M^2_P= M^3_* V_2$, we get $\lambda_{3p}\sim \frac{T_3}{M^4_*}\ll 1$ as required for the negligible backreaction on the volume modulus. Therefore, for $|v_p|\sim M_*^5$, we need a small violation of scaling symmetry as $|1-s|\ll 1$. Thus, for $|1-s|\ll 1$, $\lambda_{1p}\simeq \lambda_{2p}\simeq -\frac{\lambda_{3p}}{1-s}$.
Then, the conditions, $\lambda_{2p}>0$ and $2\lambda_{2p}+\lambda_{3p}>0$,
and $\lambda_{3p}>0$ for four-dimensional  de-Sitter space, lead to $1-s<0$ and $v_p<0$. In this case, the effective brane tension,
$T_3=\frac{\pi}{2}(1-s)v_p R_0$ is positive.

For these choices of $\la_{1p,2p,3p}$, since $\phi_f=4\ln W_f$, we can expand the effective potential of the probe brane (\ref{nonconeffa}) for $|1-s|\ll 1$ as
\be
V_{\rm eff}\simeq  T_3 W^4_f \Big[1+(1-s)\ln W_f \Big].
\ee
As a result, for a small breaking of scaling invariance, the effective action approach for a non-conical probe provides the same potential with the DBI action for a conical probe. Furthermore, the effective 3-brane tension is naturally suppressed by a factor of $(1-s)$ compared to the six dimensional  fundamental scale (assuming that naturally $v_p^{1/5}\sim R_0^{-1} \sim M_*$).

\def\theequation{D.\arabic{equation}}
\setcounter{equation}{0}

\vskip0.8cm
\noindent
{\Large \bf Appendix D: Conical multi-brane warped solutions}
\vskip0.4cm
\noindent

The general multi-brane warped solution with conical branes \cite{multibranes} admits only four-dimensional Minkowski spacetime
and takes the following form in complex internal space coordinates\footnote{Compared to Ref.~\cite{multibranes}, we changed the solution form to be compatible with the notations in our paper. For instance, the dilaton $\phi_{LL}$ is changed to $-\frac{1}{2}\phi$ and the dilaton constant is set to zero. The convention for the six-dimensional fundamental scale in \cite{multibranes} is $M^4_{*LL}=1$, but we restore $M_*$ when necessary.},
\bea
ds^2&=&{\hat W}^2\eta_{\mu\nu}dx^\mu dx^\nu +\frac{1}{2|V(z)|^2}\frac{P({\hat W})}{{\hat W}^2}dz d{\bar z}~,\\
F_{mn}&=&f e^{-\frac{1}{2}\phi}{\hat W}^{-4}\epsilon_{mn}~, \\
\phi&=& 4\ln {\hat W}~,
\eea
where $f$ is a constant parameter, $V(z)$ is an arbitrary holomorphic function and the function $P({\hat W})$ is given by
\be
P({\hat W})=\frac{1}{8}g^2{\hat W}^{-4}(W^4_+-{\hat W}^4)({\hat W}^4-W^4_-)~.
\ee
The hatted warped factor ${\hat W}$ is determined implicitly by the following algebraic equation
\be
\frac{({\hat W}^4(\zeta)-W^4_-)^{W^4_-}}{(W^4_+-{\hat W}^4(\zeta))^{W^4_+}}={\rm exp}\Big\{\frac{1}{2}g^2(W^4_+-W^4_-)(\zeta-\zeta_0)\Big\}~,\label{warpfactor}
\ee
where $\zeta_0$ is an integration constant of the warp factor,
\be
\zeta(z)=\frac{1}{2}\Big(\int^z\frac{d\omega}{V(\omega)}+{\rm c.c.}\Big)~, \label{zeta}
\ee
and
\be
W^4_\pm =v\pm \sqrt{v^2-\frac{f^2}{4g^2}}~, \label{wpm}
\ee
with $v$ being an integration constant.
Here, we note that there exists a conical singularity with nontrivial deficit angle or nontrivial brane tension
only at $\zeta=\pm\infty$ or ${\hat W}=W_\pm$. The warp factor lies in the finite range, $W_-\leq {\hat W}\leq W_+$.

Let us  now review  the matching conditions at the conical singularities depending on the holomorphic function $V(z)$ \cite{multibranes}.
Suppose that the holomorphic function is expanded around $z_i$ as $V\sim \frac{1}{c_i}(z-z_i)^{\alpha_i}$.
Then, for integer $\alpha_i\neq 1$, $\zeta$ or the warp factor are single-valued because $\oint dz/z^{\alpha_i}=0$. On the other hand,
 for $\alpha_i=1$, $\zeta$ or the warp factor are single-valued only for real $c_i$ because $\oint dz/z=2\pi i$.
When $\alpha_i$ is non-integer, there is a branch cut of $V(z)$ along ${\rm Re}(z-z_i)>0$ and therefore there is a line-like singularity along the branch cut, unless the space is covered with multiple patches. Moreover, for $\alpha_i<-1$, there is a curvature singularity \cite{multibranes}.
At a simple zero with $\alpha_i=1$, there appears a conical singularity at $z_i$ and
the matching condition for a brane tension is \cite{multibranes}
\bea
c_i>0&:& ~~T_{i+}=2\pi M^4_*\Big[1-\frac{1}{4}|c_i|g^2\Big(\frac{W^4_+}{W^4_-}-1\Big)\Big]~, \\
c_i<0&:& ~~T_{i-}=2\pi M^4_*\Big[1-\frac{1}{4}|c_i|g^2\Big(1-\frac{W^4_-}{W^4_+}\Big)\Big]~.
\eea
At a simple pole with $\alpha_i=-1$, one has to introduce a fixed negative tension
with $T_i=-2\pi M^4_*$ at $z_i$.

For instance, when $V(z)=\frac{z}{|c|}$ with $|c|$ being a constant, the warped solution takes the following form\cite{multibranes},
\bea
ds^2={\hat W}^2\eta_{\mu\nu}dx^\mu dx^\nu +\frac{{\hat W}^4}{2P({\hat W})}d{\hat W}^2+|c|^2\frac{P({\hat W})}{2{\hat W}^2}d\theta^2~,
\eea
where use is made of $\frac{dz}{V(z)}=d\zeta+i|c|d\theta$, $\frac{d{\bar z}}{{\overline V(z)}}=d\zeta-i|c|d\theta$ with $d\zeta=\frac{{\hat W}^3}{P({\hat W})}d{\hat W}$, and the angular coordinate $\theta$ ranges from $0$ to $2\pi$.
Then, by making a coordinate transformation, ${\hat W}^2=W^2_- W^2(r)$, comparing the parameters as $\frac{W^4_+}{W^4_-}=\frac{r^2_0}{r^2_1}$, $f=W^4_- q$ and
$\lambda=\frac{1}{4}|c|g^2\Big(\frac{W^4_+}{W^4_-}-1\Big)$, and finally making a scaling transformation, $g_{MN}\rightarrow e^{\frac{1}{2}\phi_0-2\ln W_-}g_{MN}$ and $\phi\rightarrow \phi+\phi_0-4\ln W^4_-$,
we reach the same form of the warped solution with two conical branes as in eq.~(\ref{wmetric}).
In the complex coordinate, we recover that one brane with tension $T_1=2\pi M_*^4(1-\lambda)$ is located at $z=0$
and the other brane with tension $T_2=2\pi M_*^4\Big(1-\lambda\frac{r^2_1}{r^2_0}
\Big)$ is located at $|z|=\infty$.

When $V(z)$ has multiple zeros, it is possible to accommodate multiple branes\cite{multibranes}.
In order for the additional brane(s) not to change the brane junction conditions of the two branes of the previous solution, we need to impose that the holomorphic function has the limit $V(z)\rightarrow\frac{z}{|c|}$ for $z\rightarrow 0$ and $|z|\rightarrow \infty$.
The simple example that satisfies these condition is
\be
V(z)=\frac{z}{|c|}\Big(1+\frac{\alpha}{z+z^{-1}}\Big)~, \label{multibrane}
\ee
where $\alpha$ is a constant parameter.
From eq.~(\ref{zeta}), we obtain the $\zeta$ variable as $\zeta=|c|\ln|z|^2+|c|(-a\ln(z-z_+)+a\ln(z-z_-)+{\rm c.c.})$
with $a=\frac{\alpha}{\sqrt{\alpha^2-4}}$. Therefore, from eq.~(\ref{warpfactor}), the warp factor behavior is determined by the simple zeros, $z=0$ and $z=z_\pm$. However, the positions of simple poles, $z=\pm i$, do not affect the warp factor change.

For single-valued variable $\zeta$ or warp factor, $\alpha$ must be a real number with $|\alpha|>2$ or a pure imaginary.
Then, this holomorphic function has three simple zeros at $z=0,z_\pm=\frac{1}{2}(-\alpha\pm\sqrt{\alpha^2-4})$ at which $\zeta$ diverges and it has two simple poles at $z=\pm i$.
It has been shown in \cite{multibranes} that one must place an arbitrary tension brane at a simple zero and a fixed tension brane with $T_{\pm i}=-2\pi M^4_*$ at a simple pole.
Therefore, on top of two branes given in eqs.~(\ref{tension1}) and (\ref{tension2}), there are two additional branes with nontrivial tension possible at $z=z_\pm$.
For a real $\alpha$, there are two possibilities:
for $\alpha<-2$,
\bea
T_3&=&2\pi M^4_* (1-\lambda |a|)~, \quad {\rm at} \ z=z_+,  \label{t3}\\
T_4&=&2\pi M^4_* \Big(1-\lambda |a| \frac{r^2_1}{r^2_0}\Big)~, \quad {\rm at} \ z=z_-; \label{t4}
\eea
for $\alpha>2$,
\bea
T_3&=&2\pi M^4_* (1-\lambda |a|)~, \quad {\rm at} \ z=z_-,  \label{t3a}\\
T_4&=&2\pi M^4_* \Big(1-\lambda |a| \frac{r^2_1}{r^2_0}\Big)~, \quad {\rm at} \ z=z_+. \label{t4a}
\eea
On the other hand, for a pure imaginary $\alpha=i\beta$ with $\beta$ a real number, we get the following additional branes:
for $\beta<0$,
\bea
T_3&=&2\pi M^4_* (1-\lambda |b|)~, \quad {\rm at} \ z=z_+,  \label{tt3}\\
T_4&=&2\pi M^4_* \Big(1-\lambda |a| \frac{r^2_1}{r^2_0}\Big)~, \quad {\rm at} \ z=z_-; \label{tt4}
\eea
for $\beta>0$,
\bea
T_3&=&2\pi M^4_* (1-\lambda |b|)~, \quad {\rm at} \ z=z_-,  \label{tt3a}\\
T_4&=&2\pi M^4_* \Big(1-\lambda |b| \frac{r^2_1}{r^2_0}\Big)~, \quad {\rm at} \ z=z_+, \label{tt4a}
\eea
where $|b|=\frac{|\beta|}{\sqrt{\beta^2+4}}$.
For $\lambda$ determined by the brane tension $T_1$ given in eq.~(\ref{tension1}), we use eq.~(\ref{t3})
to determine the parameter $a,b$ or the brane positions $z_\pm$.
The case with $\alpha=i\beta$ and $\beta>0$ was considered in section 6.3.1.

Let us finally consider the flux quantization condition for the above multi-brane solutions. Let us assume that one of the background branes is non-BPS such that the localized effective FI term differs from the supersymmetric condition, \ie $\xi_{1,{\rm eff}}= \frac{T_1}{4g}(2 / M_*^4)^{1/2}+r_1g (M_*^4/2)^{1/2}|\langle Q_1\rangle|^2\neq \xi_1$ while $\xi_{2,{\rm eff}}=\frac{T_2}{4g}(2/M_*^4)^{1/2}$. This is possible due to a nonzero VEV of the brane scalar field. On the other hand, the additional branes are non-BPS too, such that their effective FI terms can be made negligible, as assumed in the previous discussions for the probe brane.
Then, the gauge field strength (\ref{gaugefieldst}) becomes
\be
\langle F_{mn}\rangle =\langle {\hat F}_{mn}\rangle +\sum_{i=1,2} \left({2 \over  M_*^4} \right)^{1/2}\xi_{i,{\rm eff}}\,\epsilon_{mn}\frac{\delta^2(y-y_i)}{e_2}~,
\ee
where $\langle {\hat F}_{mn}\rangle$ is a regular piece of the background solution.
From the flux quantization condition
\be
g\int_{{\cal M}_2} \langle F_2\rangle = 2\pi n~,
\ee
we find, as in (\ref{quant}), the correction terms coming from the localized FI terms as following
\be
g\int_{{\cal M}_2} \langle {\hat F}_2\rangle = 2\pi \left[n-\left({2 \over  M_*^4} \right)^{1/2}\frac{g}{2\pi}\sum_{i=1,2}\xi_{i,{\rm eff}}\right]~.
\ee
In the particular example with $V(z)$ as in eq.~(\ref{multibrane}), we obtain the above relation between the tensions of the various branes
\be
\Big(2-\frac{T_1+T_3}{2\pi M_*^4}\Big)\Big(2-\frac{T_2+T_4}{2\pi M_*^4}\Big)
= \bigg[n-\frac{1}{4\pi M_*^4}\Big(T_1+2r_1g^2 M_*^4|\langle Q_1\rangle|^2+ T_2\Big)\bigg]^2~.
\ee
The above condition relating the brane tensions is to be compared with (\ref{quant2}) for the two background brane case. The different numerical factors in the left hand side have to do with the details of the construction of this particular solution.


\begin{thebibliography}{999}



\bibitem{inflation}
%\cite{Guth:1980zm}
%\bibitem{Guth:1980zm}
  A.~H.~Guth,
  %``The Inflationary Universe: A Possible Solution To The Horizon And Flatness
  %Problems,''
  Phys.\ Rev.\  D {\bf 23} (1981) 347;
  %%CITATION = PHRVA,D23,347;%%
%\cite{Linde:1981mu}
%\bibitem{Linde:1981mu}
  A.~D.~Linde,
  %``A New Inflationary Universe Scenario: A Possible Solution Of The Horizon,
  %Flatness, Homogeneity, Isotropy And Primordial Monopole Problems,''
  Phys.\ Lett.\  B {\bf 108} (1982) 389;
  %%CITATION = PHLTA,B108,389;%%
%\cite{Albrecht:1982wi}
%\bibitem{Albrecht:1982wi}
  A.~Albrecht and P.~J.~Steinhardt,
  %``Cosmology For Grand Unified Theories With Radiatively Induced Symmetry
  %Breaking,''
  Phys.\ Rev.\ Lett.\  {\bf 48} (1982) 1220;
  %%CITATION = PRLTA,48,1220;%%
%\cite{Linde:1983gd}
%\bibitem{Linde:1983gd}
  A.~D.~Linde,
  %``Chaotic Inflation,''
  Phys.\ Lett.\  B {\bf 129} (1983) 177.
  %%CITATION = PHLTA,B129,177;%%








\bibitem{reviews}
%\cite{Quevedo:2002xw}
%\bibitem{Quevedo:2002xw}
  F.~Quevedo,
  %``Lectures on string/brane cosmology,''
  Class.\ Quant.\ Grav.\  {\bf 19} (2002) 5721
  [arXiv:hep-th/0210292];
  %%CITATION = CQGRD,19,5721;%%
%\cite{Linde:2005dd}
%\bibitem{Linde:2005dd}
  A.~Linde,
  %``Inflation and string cosmology,''
  eConf {\bf C040802} (2004) L024
  [J.\ Phys.\ Conf.\ Ser.\  {\bf 24} (2005\ PTPSA,163,295-322.2006) 151]
  [arXiv:hep-th/0503195];
  %%CITATION = PTPSA,163,295;%%
 %\cite{HenryTye:2006uv}
%\bibitem{HenryTye:2006uv}
  S.~H.~Henry Tye,
  %``Brane inflation: String theory viewed from the cosmos,''
  Lect.\ Notes Phys.\  {\bf 737} (2008) 949
  [arXiv:hep-th/0610221];
  %%CITATION = LNPHA,737,949;%%
%\cite{Cline:2006hu}
%\bibitem{Cline:2006hu}
  J.~M.~Cline,
  %``String cosmology,''
  arXiv:hep-th/0612129;
  %%CITATION = HEP-TH/0612129;%%
 R.~Kallosh,
  %``On Inflation in String Theory,''
  Lect.\ Notes Phys.\  {\bf 738} (2008) 119
  [arXiv:hep-th/0702059];
  %%CITATION = LNPHA,738,119;%%
%\cite{Burgess:2007pz}
%\bibitem{Burgess:2007pz}
  C.~P.~Burgess,
  %``Lectures on Cosmic Inflation and its Potential Stringy Realizations,''
  PoS {\bf P2GC} (2006) 008
  [Class.\ Quant.\ Grav.\  {\bf 24} (2007) S795]
  [arXiv:0708.2865 [hep-th]];
  %%CITATION = CQGRD,24,S795;%%
 %\cite{McAllister:2007bg}
%\bibitem{McAllister:2007bg}
  L.~McAllister and E.~Silverstein,
  %``String Cosmology: A Review,''
  Gen.\ Rel.\ Grav.\  {\bf 40} (2008) 565
  [arXiv:0710.2951 [hep-th]].
  %%CITATION = GRGVA,40,565;%%




\bibitem{KKLT}
  %\cite{Kachru:2003sx}
%\bibitem{Kachru:2003sx}
  S.~Kachru, R.~Kallosh, A.~Linde, J.~M.~Maldacena, L.~P.~McAllister and S.~P.~Trivedi,
  %``Towards inflation in string theory,''
  JCAP {\bf 0310} (2003) 013
  [arXiv:hep-th/0308055].
  %%CITATION = JCAPA,0310,013;%%




\bibitem{DT}
  G.~R.~Dvali and S.~H.~H.~Tye,
  %``Brane inflation,''
  Phys.\ Lett.\  B {\bf 450} (1999) 72
  [arXiv:hep-ph/9812483];
  %%CITATION = PHLTA,B450,72;%%
  %\cite{Burgess:2001fx}
%\bibitem{Burgess:2001fx}
  C.~P.~Burgess, M.~Majumdar, D.~Nolte, F.~Quevedo, G.~Rajesh and R.~J.~Zhang,
  %``The Inflationary Brane-Antibrane Universe,''
  JHEP {\bf 0107} (2001) 047
  [arXiv:hep-th/0105204];
  %%CITATION = JHEPA,0107,047;%%
%\cite{GarciaBellido:2001ky}
%\bibitem{GarciaBellido:2001ky}
  J.~Garcia-Bellido, R.~Rabadan and F.~Zamora,
  %``Inflationary scenarios from branes at angles,''
  JHEP {\bf 0201} (2002) 036
  [arXiv:hep-th/0112147].
  %%CITATION = JHEPA,0201,036;%%






\bibitem{dbi}
%\cite{Silverstein:2003hf}
%\bibitem{Silverstein:2003hf}
  E.~Silverstein and D.~Tong,
  %``Scalar Speed Limits and Cosmology: Acceleration from D-cceleration,''
  Phys.\ Rev.\  D {\bf 70} (2004) 103505
  [arXiv:hep-th/0310221];
  %%CITATION = PHRVA,D70,103505;%%
  M.~Alishahiha, E.~Silverstein and D.~Tong,
  %``DBI in the sky,''
  Phys.\ Rev.\  D {\bf 70} (2004) 123505
  [arXiv:hep-th/0404084];
 %\cite{Chen:2005ad}
%\bibitem{Chen:2005ad}
  X.~Chen,
  %``Inflation from warped space,''
  JHEP {\bf 0508} (2005) 045
  [arXiv:hep-th/0501184];
  %%CITATION = JHEPA,0508,045;%%
%\cite{Bean:2007hc}
%\bibitem{Bean:2007hc}
  R.~Bean, S.~E.~Shandera, S.~H.~Henry Tye and J.~Xu,
  %``Comparing Brane Inflation to WMAP,''
  JCAP {\bf 0705} (2007) 004
  [arXiv:hep-th/0702107];
  %%CITATION = JCAPA,0705,004;%%
%\cite{Becker:2007ui}
%\bibitem{Becker:2007ui}
  M.~Becker, L.~Leblond and S.~E.~Shandera,
  %``Inflation from Wrapped Branes,''
  Phys.\ Rev.\  D {\bf 76} (2007) 123516
  [arXiv:0709.1170 [hep-th]].
  %%CITATION = PHRVA,D76,123516;%%


\bibitem{angularmotion1}
%\cite{Easson:2007dh}
%\bibitem{Easson:2007dh}
  D.~A.~Easson, R.~Gregory, D.~F.~Mota, G.~Tasinato and I.~Zavala,
  %``Spinflation,''
  JCAP {\bf 0802} (2008) 010
  [arXiv:0709.2666 [hep-th]].
  %%CITATION = JCAPA,0802,010;%%


\bibitem{dbihigher}
%\cite{Langlois:2008wt}
%\bibitem{Langlois:2008wt}
  D.~Langlois, S.~Renaux-Petel, D.~A.~Steer and T.~Tanaka,
  %``Primordial fluctuations and non-Gaussianities in multi-field DBI
  %inflation,''
  Phys.\ Rev.\ Lett.\  {\bf 101} (2008) 061301
  [arXiv:0804.3139 [hep-th]];
  %%CITATION = PRLTA,101,061301;%%
%\cite{Langlois:2008qf}
%\bibitem{Langlois:2008qf}
  D.~Langlois, S.~Renaux-Petel, D.~A.~Steer and T.~Tanaka,
  %``Primordial perturbations and non-Gaussianities in DBI and general
  %multi-field inflation,''
  Phys.\ Rev.\  D {\bf 78} (2008) 063523
  [arXiv:0806.0336 [hep-th]].
  %%CITATION = PHRVA,D78,063523;%%







\bibitem{mirage}
A.~Kehagias and E.~Kiritsis,
  %``Mirage cosmology,''
  JHEP {\bf 9911} (1999) 022
  [arXiv:hep-th/9910174].
  %%CITATION = JHEPA,9911,022;%%


\bibitem{miragecosmo}
C.~P.~Burgess, P.~Martineau, F.~Quevedo and R.~Rabadan,
  %``Branonium,''
  JHEP {\bf 0306} (2003) 037
  [arXiv:hep-th/0303170];
  %%CITATION = JHEPA,0306,037;%%
C.~Germani, N.~E.~Grandi and A.~Kehagias,
  %``A stringy alternative to inflation: The cosmological slingshot scenario,''
  Class.\ Quant.\ Grav.\  {\bf 25} (2008) 135004
  [arXiv:hep-th/0611246].
  %%CITATION = CQGRD,25,135004;%%





\bibitem{ss}
H.~Nishino and E.~Sezgin,
  %``Matter And Gauge Couplings Of N=2 Supergravity In Six-Dimensions,''
  Phys.\ Lett.\  B {\bf 144} (1984) 187;
  %%CITATION = PHLTA,B144,187;%%
  A.~Salam and E.~Sezgin,
  %``Chiral Compactification On Minkowski X S**2 Of N=2 Einstein-Maxwell
  %Supergravity In Six-Dimensions,''
  Phys.\ Lett.\ B {\bf 147} (1984) 47.
  %%CITATION = PHLTA,B147,47;%%

\bibitem{cline}
 F.~Chen, J.~M.~Cline and S.~Kanno,
  %``Modified Friedmann Equation and Inflation in Warped Codimension-two
  %Braneworld,''
  Phys.\ Rev.\  D {\bf 77} (2008) 063531
  [arXiv:0801.0226 [hep-th]].
  %%CITATION = PHRVA,D77,063531;%%






\bibitem{6dself}
 S.~M.~Carroll and M.~M.~Guica,
%``Sidestepping the cosmological constant with football-shaped extra
%dimensions,''
arXiv:hep-th/0302067;
%%CITATION = HEP-TH 0302067;%%
I.~Navarro,
%``Codimension two compactifications and the cosmological constant  problem,''
JCAP {\bf 0309} (2003) 004 [arXiv:hep-th/0302129];
%\bibitem{football}
  Y.~Aghababaie, C.~P.~Burgess, S.~L.~Parameswaran and F.~Quevedo,
  %``Towards a naturally small cosmological constant from branes in 6D
  %supergravity,''
  Nucl.\ Phys.\ B {\bf 680} (2004) 389
  [arXiv:hep-th/0304256];
  %%CITATION = HEP-TH 0304256;%%
I.~Navarro,
  %``Spheres, deficit angles and the cosmological constant,''
  Class.\ Quant.\ Grav.\  {\bf 20} (2003) 3603
  [arXiv:hep-th/0305014];
  %%CITATION = CQGRD,20,3603;%%
  H.~P.~Nilles, A.~Papazoglou and G.~Tasinato,
%``Selftuning and its footprints,''
Nucl.\ Phys.\ B {\bf 677} (2004) 405
[arXiv:hep-th/0309042];
%%CITATION = HEP-TH 0309042;%%
%\cite{Lee:2003wg}
%\bibitem{Lee:2003wg}
  H.~M.~Lee,
  %``A comment on the self-tuning of cosmological constant with deficit  angle
  %on a sphere,''
  Phys.\ Lett.\  B {\bf 587} (2004) 117
  [arXiv:hep-th/0309050];
  %%CITATION = PHLTA,B587,117;%%
  A.~Kehagias,
  %``A conical tear drop as a vacuum-energy drain for the solution of the
  %cosmological constant problem,''
  Phys.\ Lett.\  B {\bf 600} (2004) 133
  [arXiv:hep-th/0406025];
  %%CITATION = PHLTA,B600,133;%%
S.~Randjbar-Daemi and V.~A.~Rubakov,
  %``4d-flat compactifications with brane vorticities,''
  JHEP {\bf 0410} (2004) 054
  [arXiv:hep-th/0407176];
  %%CITATION = JHEPA,0410,054;%%
H.~M.~Lee and A.~Papazoglou,
  %``Brane solutions of a spherical sigma model in six dimensions,''
  Nucl.\ Phys.\  B {\bf 705} (2005) 152
  [arXiv:hep-th/0407208];
  %%CITATION = NUPHA,B705,152;%%
  A.~Kehagias and C.~Mattheopoulou,
  %``Flat-brane Compactifications in Supergravity Induced by Scalars,''
  Nucl.\ Phys.\  B {\bf 797} (2008) 117
  [arXiv:0710.4021 [hep-th]];
  %%CITATION = NUPHA,B797,117;%%
 K.~Koyama,
  %``The cosmological constant and dark energy in braneworlds,''
  Gen.\ Rel.\ Grav.\  {\bf 40} (2008) 421
  [arXiv:0706.1557 [astro-ph]];
  %%CITATION = GRGVA,40,421;%%
%\cite{Burgess:2008yx}
%\bibitem{Burgess:2008yx}
  C.~P.~Burgess, D.~Hoover, C.~de Rham and G.~Tasinato,
  %``Effective Field Theories and Matching for Codimension-2 Branes,''
  arXiv:0812.3820 [hep-th].
  %%CITATION = ARXIV:0812.3820;%%



\bibitem{6dcosmo}
J.~Vinet and J.~M.~Cline,
  %``Can codimension-two branes solve the cosmological constant problem?,''
  Phys.\ Rev.\ D {\bf 70} (2004) 083514
  [arXiv:hep-th/0406141];
  %%CITATION = HEP-TH 0406141;%%
J.~Vinet and J.~M.~Cline,
  %``Codimension-two branes in six-dimensional supergravity and the
  %cosmological constant problem,''
  Phys.\ Rev.\ D {\bf 71} (2005) 064011
  [arXiv:hep-th/0501098];
  %%CITATION = HEP-TH 0501098;%%
A.~J.~Tolley, C.~P.~Burgess, C.~de Rham and D.~Hoover,
  %``Scaling solutions to 6D gauged chiral supergravity,''
  New J.\ Phys.\  {\bf 8} (2006) 324
  [arXiv:hep-th/0608083];
  %%CITATION = NJOPF,8,324;%%
B.~Himmetoglu and M.~Peloso,
  %``Isolated Minkowski vacua, and stability analysis for an extended brane in
  %the rugby ball,''
  Nucl.\ Phys.\  B {\bf 773} (2007) 84
  [arXiv:hep-th/0612140];
  %%CITATION = NUPHA,B773,84;%%
T.~Kobayashi and M.~Minamitsuji,
  %``Brane cosmological solutions in six-dimensional warped flux
  %compactifications,''
  JCAP {\bf 0707} (2007) 016
  [arXiv:0705.3500 [hep-th]];
  %%CITATION = JCAPA,0707,016;%%
E.~J.~Copeland and O.~Seto,
  %``Dynamical solutions of warped six dimensional supergravity,''
  JHEP {\bf 0708} (2007) 001
  [arXiv:0705.4169 [hep-th]];
  %%CITATION = JHEPA,0708,001;%%
E.~Papantonopoulos, A.~Papazoglou and V.~Zamarias,
  %``Induced cosmology on a regularized brane in six-dimensional flux
  %compactification,''
  Nucl.\ Phys.\  B {\bf 797} (2008) 520
  [arXiv:0707.1396 [hep-th]];
  %%CITATION = NUPHA,B797,520;%%
 M.~Minamitsuji and D.~Langlois,
  %``Cosmological evolution of regularized branes in 6D warped flux
  %compactifications,''
  Phys.\ Rev.\  D {\bf 76} (2007) 084031
  [arXiv:0707.1426 [hep-th]].
  %%CITATION = PHRVA,D76,084031;%%


\bibitem{GBind}
P.~Bostock, R.~Gregory, I.~Navarro and J.~Santiago,
  %``Einstein gravity on the codimension 2 brane?,''
  Phys.\ Rev.\ Lett.\  {\bf 92} (2004) 221601
  [arXiv:hep-th/0311074];
  %%CITATION = PRLTA,92,221601;%%
E.~Papantonopoulos and A.~Papazoglou,
  %``Brane-bulk matter relation for a purely conical codimension-2 brane
  %world,''
  JCAP {\bf 0507}, 004 (2005)
  [arXiv:hep-th/0501112];
  %%CITATION = JCAPA,0507,004;%%
N.~Kaloper and D.~Kiley,
  %``Exact black holes and gravitational shockwaves on codimension-2 branes,''
  JHEP {\bf 0603}, 077 (2006)
  [arXiv:hep-th/0601110].
  %%CITATION = JHEPA,0603,077;%%
N.~Kaloper,
  %``Brane Induced Gravity: Codimension-2,''
  Mod.\ Phys.\ Lett.\  A {\bf 23}, 781 (2008)
  [arXiv:0711.3210 [hep-th]].
  %%CITATION = MPLAE,A23,781;%%
C.~Charmousis and A.~Papazoglou,
  %``Self-properties of codimension-2 braneworlds,''
  JHEP {\bf 0807} (2008) 062
  [arXiv:0804.2121 [hep-th]];
  %%CITATION = JHEPA,0807,062;%%
 B.~Cuadros-Melgar, E.~Papantonopoulos, M.~Tsoukalas and V.~Zamarias,
  %``Black Holes on Thin 3-branes of Codimension-2 and their Extension into the
  %Bulk,''
  Nucl.\ Phys.\  B {\bf 810}, 246 (2009)
  [arXiv:0804.4459 [hep-th]].
  %%CITATION = NUPHA,B810,246;%%












\bibitem{leepapa}
%\cite{Lee:2007vy}
%\bibitem{Lee:2007vy}
  H.~M.~Lee and A.~Papazoglou,
  %``Supersymmetric codimension-two branes in six-dimensional gauged
  %supergravity,''
  JHEP {\bf 0801} (2008) 008
  [arXiv:0710.4319 [hep-th]].
  %%CITATION = JHEPA,0801,008;%%


\bibitem{lee}
%\cite{Lee:2008zs}
%\bibitem{Lee:2008zs}
  H.~M.~Lee,
  %``Supersymmetric codimension-two branes and U(1)_R mediation in 6D gauged
  %supergravity,''
  JHEP {\bf 0805} (2008) 028
  [arXiv:0803.2683 [hep-th]].
  %%CITATION = JHEPA,0805,028;%%







\bibitem{choilee}
%\cite{Choi:2009jd}
%\bibitem{Choi:2009jd}
  K.~Y.~Choi and H.~M.~Lee,
  %``U(1)_R-mediated supersymmetry breaking from a six-dimensional flux
  %compactification,''
  arXiv:0901.3545 [hep-ph].
  %%CITATION = ARXIV:0901.3545;%%




\bibitem{uvcap}
%\cite{Burgess:2007vi}
%\bibitem{Burgess:2007vi}
  C.~P.~Burgess, D.~Hoover and G.~Tasinato,
  %``UV Caps and Modulus Stabilization for 6D Gauged Chiral Supergravity,''
  JHEP {\bf 0709} (2007) 124
  [arXiv:0705.3212 [hep-th]].
  %%CITATION = JHEPA,0709,124;%%


\bibitem{regular}
J.~R.~I.~Gott,
  %``Gravitational lensing effects of vacuum strings: Exact solutions,''
  Astrophys.\ J.\  {\bf 288} (1985) 422;
  %%CITATION = ASJOA,288,422;%%
   E.~Papantonopoulos, A.~Papazoglou and V.~Zamarias,
   %``Regularization of conical singularities in warped six-dimensional
   %compactifications,''
   JHEP {\bf 0703} (2007) 002
   [arXiv:hep-th/0611311];
   %%CITATION = JHEPA,0703,002;%%
N.~Kaloper and D.~Kiley,
  %``Charting the Landscape of Modified Gravity,''
  JHEP {\bf 0705} (2007) 045
  [arXiv:hep-th/0703190];
  %%CITATION = JHEPA,0705,045;%%
C.~de Rham,
  %``The Effective Field Theory of Codimension-two Branes,''
  JHEP {\bf 0801} (2008) 060
  [arXiv:0707.0884 [hep-th]];
  %%CITATION = JHEPA,0801,060;%%
T.~Kobayashi, T.~Shiromizu and C.~de Rham,
  %``Curvature corrections to the low energy effective theory in 6D regularized
  %braneworlds,''
  Phys.\ Rev.\  D {\bf 77} (2008) 124012
  [arXiv:0802.0103 [hep-th]].
  %%CITATION = PHRVA,D77,124012;%%



\bibitem{4dgravity}
M.~Peloso, L.~Sorbo and G.~Tasinato,
   %``Standard 4d gravity on a brane in six dimensional flux
%compactifications,''
   Phys.\ Rev.\  D {\bf 73} (2006) 104025
   [arXiv:hep-th/0603026];
   %%CITATION = PHRVA,D73,104025;%%
T.~Kobayashi and M.~Minamitsuji,
  %``Gravity on an extended brane in six-dimensional warped flux
  %compactifications,''
  Phys.\ Rev.\  D {\bf 75} (2007) 104013
  [arXiv:hep-th/0703029].
  %%CITATION = PHRVA,D75,104013;%%
\bibitem{clown}
J.~M.~Cline, J.~Descheneau, M.~Giovannini and J.~Vinet,
  %``Cosmology of codimension-two braneworlds,''
  JHEP {\bf 0306} (2003) 048
  [arXiv:hep-th/0304147].
  %%CITATION = JHEPA,0306,048;%%




\bibitem{hybrid}
%\cite{Linde:1991km}
%\bibitem{Linde:1991km}
  A.~D.~Linde,
  %``Axions in inflationary cosmology,''
  Phys.\ Lett.\  B {\bf 259} (1991) 38.
  %%CITATION = PHLTA,B259,38;%%




\bibitem{gibbons}
G.~W.~Gibbons, R.~Guven and C.~N.~Pope,
  %``3-branes and uniqueness of the Salam-Sezgin vacuum,''
  Phys.\ Lett.\ B {\bf 595} (2004) 498
  [arXiv:hep-th/0307238].
  %%CITATION = HEP-TH 0307238;%%


\bibitem{burgess03}
Y.~Aghababaie {\it et al.},
  %``Warped brane worlds in six dimensional supergravity,''
  JHEP {\bf 0309} (2003) 037
  [arXiv:hep-th/0308064].
    %%CITATION = HEP-TH 0308064;%%


\bibitem{multibranes}
%\cite{Lee:2005az}
%\bibitem{Lee:2005az}
  H.~M.~Lee and C.~L\"udeling,
  %``The general warped solution with conical branes in six-dimensional
  %supergravity,''
  JHEP {\bf 0601} (2006) 062
  [arXiv:hep-th/0510026].
  %%CITATION = JHEPA,0601,062;%%




\bibitem{leepapa1}
H.~M.~Lee and A.~Papazoglou, work in progress.



\bibitem{quevedo}
%\cite{Aghababaie:2003wz}
%\bibitem{Aghababaie:2003wz}
  Y.~Aghababaie, C.~P.~Burgess, S.~L.~Parameswaran and F.~Quevedo,
  %``Towards a naturally small cosmological constant from branes in 6D
  %supergravity,''
  Nucl.\ Phys.\  B {\bf 680} (2004) 389
  [arXiv:hep-th/0304256].
  %%CITATION = NUPHA,B680,389;%%









\bibitem{riottolyth}
%\cite{Lyth:1998xn}
%\bibitem{Lyth:1998xn}
  D.~H.~Lyth and A.~Riotto,
  %``Particle physics models of inflation and the cosmological density
  %perturbation,''
  Phys.\ Rept.\  {\bf 314} (1999) 1
  [arXiv:hep-ph/9807278].
  %%CITATION = PRPLC,314,1;%%


\bibitem{wmap5}
%\cite{Komatsu:2008hk}
%\bibitem{Komatsu:2008hk}
  E.~Komatsu {\it et al.}  [WMAP Collaboration],
  %``Five-Year Wilkinson Microwave Anisotropy Probe (WMAP\altaffilmark 1 )
  %Observations:Cosmological Interpretation,''
  arXiv:0803.0547 [astro-ph].
  %%CITATION = ARXIV:0803.0547;%%



\bibitem{CLP}
  J.~W.~Chen, M.~A.~Luty and E.~Ponton,
  %``A critical cosmological constant from millimeter extra dimensions,''
  JHEP {\bf 0009} (2000) 012
  [arXiv:hep-th/0003067].
  %%CITATION = JHEPA,0009,012;%%





\bibitem{curved}
%\cite{Tolley:2005nu}
%\bibitem{Tolley:2005nu}
  A.~J.~Tolley, C.~P.~Burgess, D.~Hoover and Y.~Aghababaie,
  %``Bulk singularities and the effective cosmological constant for higher
  %co-dimension branes,''
  JHEP {\bf 0603} (2006) 091
  [arXiv:hep-th/0512218].
  %%CITATION = JHEPA,0603,091;%%








\end{thebibliography}
\end{document}